\documentclass[12pt,preprint]{aastex}

\usepackage{epsfig}

\slugcomment{In Revision for the Astrophysical Journal}

\shorttitle{Residence Times of Dust in Protoplanetary Disk Environments}
\shortauthors{F. J. Ciesla}

\begin{document}


\title{Residence Times of Particles in Diffusive Protoplanetary Disk Environments II. Radial Motions and Applications to Dust Annealing}

\author{F. J. Ciesla\altaffilmark{1}}
\affil{Department of the Geophysical Sciences, The University of Chicago, 5734 South Ellis Avenue, Chicago, IL 60637}
\newpage

\begin{abstract}

The origin of crystalline grains in comets and the outer regions of protoplanetary disks remains a mystery.  It has been suggested that such grains form via annealing of amorphous precursors in the hot, inner region of a protoplanetary disk, where the temperatures needed for such transformations were found, and were then transported outward by some dynamical means.  Here we develop a means of tracking the paths that dust grains would have taken through a diffusive protoplanetary disk and examine the types and ranges of environments that particles  would have seen over a 10$^{6}$ year time period in the dynamic disk.    We then combine this model with three annealing laws to examine how the dynamic evolution of amorphous grains would have led to their physical restructuring and their delivery to various regions of the disk. It is found that ``sibling particles''-- those particles that reside at the same location at a given period of time--take a wide range of unique and independent paths through the disk to arrive there.  While high temperatures can persist in the disk for very long time periods, we find that those grains which are delivered to the cold outer regions of the disk are largely annealed in the first few $\times$10$^{5}$ yrs of disk history.  This suggests that the crystallinity of grains in the outer disk would be determined early and remain unchanged for much of disk history, in agreement with recent astronomical observations.

\end{abstract}


\newpage

\section{Introduction}

Evidence for the large-scale redistribution of materials during the earliest stages of planet formation is found both in primitive objects in the Solar System as well as astronomical observations of protoplanetary disks. The STARDUST spacecraft returned crystalline silicates and refractory grains which are chemically and isotopically similar to those materials found in chondritic meteorites, pointing to a common origin \citep{brownlee06,mckeegan06,zolensky06}.  As such, it is thought that these high-temperature products formed in the solar nebula close to the Sun, where the temperatures needed for those objects to form ($> \sim$1000 K) were naturally expected, and were subsequently carried outwards by some dynamic process.  Given that crystalline silicates are observed in the cold, outer regions of disks around other stars \citep[e.g.][]{vanboekel04,apai05,watson09,sargent09,sicilia09,oliveira11}--environments similar to that in which our comets are thought to have formed--it is thought that the same dynamical processes are also responsible and are natural consequences of protoplanetary disk evolution.  

A number of mechanisms have been suggested as driving this redistribution of materials in protoplanetary disks, including turbulent diffusion \citep[e.g.][]{gail01,bockelee02,cuzzi03, ciesla10cai}, spiral arms in gravitationally unstable disks \citep{boss08}, photophoresis \citep{mousis07}, large-scale flows associated with mass and angular momentum transport \citep{kg04, ciesla07,ciesla09,ciesla10cai,desch07,jacquet11},  bipolar outflows and jets \citep[e.g][]{shu96}, and radiation pressure \citep{vinkovic09}.  Each of these processes has been shown to carry materials from the hot, inner regions of a protoplanetary disk to the cooler, outer regions if appropriate conditions are present in a protoplanetary disk.  However, it remains unclear whether these various conditions were met and thus which of these was primarily responsible for the large-scale redistribution of materials in our solar nebula.

Regardless of the exact cause of this redistribution, the fact that solid materials get transported from one location to another in a disk implies that they pass through a variety of chemical and physical environments prior to their incorporation into a comet, asteroid, or planetesimal.  As solids move through these environments, they would be destroyed or altered, the levels to which depend on the chemical and physical reactions that take place and the time spent in a given environment.  Thus grains are not simply products of a singular disk environment in which they originated, but rather record the integrated path and physical conditions to which they were exposed over their lifetime in a protoplanetary disk.  Understanding the chemical evolution of protoplanetary disks and the first building blocks of a planets requires understanding the types of paths that grains are able to take through a disk during the early evolution of a planetary system.  As the fingerprints of such environments may be found in primitive materials in comets and meteorites, this understanding may help us determine which one or ones of the dynamic processes identified above operated in the solar nebula.

The motivation of this study is to develop a method for calculating representative particle paths through a diffusive protoplanetary disk over long time periods ($>$10$^{5}$ years) and to apply that method to understand the origin of materials found in chondritic meteorites and comets.  We adopt the $\alpha$-viscosity model to describe the dynamic evolution of the disk, that is, the mass and angular momentum transport and the diffusion of species within it.  The $\alpha$-viscosity model is used because it provides a means of describing the evolution of the protoplanetary disk and the materials within it over timescales of millions of years, which is necessary when examining the data provided by meteorites and primitive materials in the Solar System.  This model assumes that the evolution of the disk is tied to some turbulent viscosity, $\nu$=$\alpha c_{s} H$, where $c_{s}$ is the local speed of sound, $H$ is the local scale-height, and $\alpha$ is some parameter ($\alpha <$1) which quantifies the strength of the turbulence of the disk.  The relation of radial diffusivity to this viscosity remains the subject of ongoing work, \citep[e.g.][]{jkm06,pavly07}, and we will begin by making the standard assumption that the diffusivity of gas in the disk is $D_{g} \sim \nu$, though the techniques described here do not require this assumption.

In the next section, the particle-tracking model is described and validated against other calculations for the radial transport of materials in a protoplanetary disk.  In Section 3 we describe the types of paths particles take through a turbulent, viscous protoplanetary disk. In Section 4 we use these results to explore how the motions of dust grains in a protoplanetary disk affect their physical structure by feeding information about the environments they saw into various kinetic models for the annealing of silicates.  In Section 5 we summarize the paper and present our conclusions.

\section{Model}

The methods used here build off of that described by \citet{ciesla10} who developed an approach to calculate representative vertical paths of particles in diffusive protoplanetary disks.   In that study, \citet{ciesla10} took the advection-diffusion equation that describes the vertical evolution of a species in a disk
\citep{dubrulle95,fromang06}:
\begin{equation}\label{vdiff}
\frac{\partial \rho_{i}}{\partial t} = \frac{\partial}{\partial z} \left( \rho_{g} D \frac{\partial \left( \frac{\rho_{i}}{\rho_{g}} \right)}{\partial z} \right) - \frac{\partial}{\partial z} \left( \rho_{i} v_{grav} \right)
\end{equation}
where $\rho_{i}$ is the spatial density of the material of interest, $\rho_{g}$ is the gas density, $D$ the diffusion coefficient, and $v_{grav}$ the vertical velocity of the species which arises due to gravitational settling and calculated the vertical motions of an individual particle using the equation:
\begin{equation}\label{vrw}
z_{i} = z_{i-1} + v_{eff} \delta t + R \left[\frac{2}{\xi} D \left(z' \right) \delta t \right]^{\frac{1}{2}}
\end{equation}
where
$z_{i-1}$ and $z_{i}$ described the position of the particle before and after a timestep of duration $\delta t$.  Eq (\ref{vrw}) came about from analyzing the Fokker-Planck equation as well as by considering the moments of Eq (\ref{vdiff}). The second term on the right of Eq. (\ref{vrw}) represents the advective motions of the particle, with the velocity not simply being the gravitational settling velocity, but includes effects that arise from the spatial variations in the gas and diffusion coefficient.  Thus,  the effective velocity, $v_{eff}$, is the sum of three terms:
\begin{equation}
v_{eff} = v_{grav} + v_{gas} + v_{FP}
\end{equation}
with $v_{grav}$ again being the vertical settling velocity set by balancing the vertical component of the gravitational force from the central star with the resistive drag force as the solid moves through the gaseous fluid, $v_{gas}$ is a term that arises due to the fact that the diffusive flux of materials depends on the ratio $\rho_{i}$/$\rho_{g}$ and $\rho_{g}$ varies with $z$ ($v_{gas}$=$\frac{D}{\rho_{g}} \frac{\partial \rho_{g}}{\partial z}$), and $v_{FP}$ is a term that accounts for possible variations in the diffusion coefficient $D$ ($v_{FP}$=$\frac{\partial D}{\partial z}$).  Detailed derivations and explanations of each term are given in \citet{ciesla10}.

The third term on the right of Eq. (\ref{vrw}) represents the displacement that occurs as a result of the diffusive motions in the disk.  This term is evaluated by finding a random number, $R$, from a distribution with variance $\xi$.  The diffusivity is evaluated at a location $z'$=$z_{i-1}$+$\frac{1}{2} \frac{\partial D}{\partial z} \delta t$ to account for strong gradients in the diffusivity.

 Here we want to develop a similar means of calculating the radial motions of materials in a diffusive protoplanetary disk as that described in \citet{ciesla10}.  The one-dimensional radial advection-diffusion equation that describes the behavior of a species, $i$, in a protoplanetary disk is written \citep[e.g.][]{gail01,bockelee02,cuzzi03,ciesla10cai}:
\begin{equation}
\frac{\partial \Sigma_{i}}{\partial t} = \frac{1}{r} \frac{\partial}{\partial r} \left( r \Sigma D \frac{\partial}{\partial r} \left( \frac{\Sigma_{i}}{\Sigma} \right) \right) - \frac{1}{r} \frac{\partial}{\partial r} \left( r v_{r} \Sigma_{i} \right)
\end{equation}
where $\Sigma_{i}$ is the surface density of the species of interest, $\Sigma$ is the surface density of the disk, $D$ is again the diffusivity, and $v_{r}$ is the radial motions of the materials due to large-scale flows and gas drag. The form of this equation is slightly different from that of Eq (\ref{vdiff}) due to the cylindrical coordinate system, and as such, leads to different, and less useful, relations between the moments of the equations.   As a result, this form of the radial transport equation does not immediately allow us to write an equation similar to Eq. (\ref{vrw}) to describe the radial trajectory of a particle in a diffusive disk.

A solution to this problem is to instead  
 adopt a Cartesian coordinate system and track the motions of the particles in the $x$-$y$ plane, where the radial distance from the origin is given by $r^{2}$=$x^{2}$+$y^{2}$.  This allows us to write the advection-diffusion equation as: 
\begin{equation}
\frac{\partial \Sigma_{i}}{\partial t} = \frac{\partial}{\partial x} \left( \Sigma D \frac{\partial}{\partial x} \left( \frac{\Sigma_{i}}{\Sigma} \right) \right) - \frac{\partial}{\partial x} \left( v_{x} \Sigma_{i} \right)
+
\frac{\partial}{\partial y} \left( \Sigma D \frac{\partial}{\partial y} \left( \frac{\Sigma_{i}}{\Sigma} \right) \right) - \frac{\partial}{\partial y} \left( v_{y} \Sigma_{i} \right)
\end{equation}
where $v_{x}$ and $v_{y}$ represent the $x$ and $y$ components of $v_{r}$.  The motions in each direction are now described using equations that have the same form as the vertical diffusion equation (Eq (\ref{vdiff})).  A similar approach is taken in other studies of transport, such as those examining the dynamics of pollutants in groundwater \citep[e.g.][]{kinzelbach90}, and has been shown to accurately describe the motions of materials of interest.

Following \citet{ciesla10},  the motions of an individual diffusing molecule or particle would be described by:
\begin{equation}\label{vrwx}
x_{i} = x_{i-1} + v_{eff,x} \delta t + R \left[\frac{2}{r} D \left(x' \right) \delta t \right]^{\frac{1}{2}}
\end{equation}
and
\begin{equation}\label{vrwy}
y_{i} = y_{i-1} + v_{eff,y} \delta t + R \left[\frac{2}{r} D \left(y' \right) \delta t \right]^{\frac{1}{2}}
\end{equation}
where each term in the above equations represents the same physical effects as described for Eq (\ref{vrw}).  Here $v_{eff,x}$ and $v_{eff,y}$  have multiple components as described above.  That is, $v_{eff,x}$=$v_{x}+v_{gas,x}+v_{FP,x}$, meaning effective velocity in the $x$-direction would be given by the sum of three different effects: the velocity of the materials in the $x$-direction due to large-scale flows and gas drag (which normally result in purely radial motions):
\begin{equation}
v_{x}\left( x_{i-1}, y_{i-1} \right) = v_{r} \frac{x_{i-1}}{r_{i-1}}
\end{equation}
where $r_{i-1}$ is the radial position of the particle at the beginning of the timestep, $v_{r}$ is the radial advective velocity evaluated at $r_{i-1}$, and $x_{i-1}$ is its position along the $x$ axis;  a contribution due to the gradient in the surface density:
\begin{equation}
v_{gas,x} \left( x_{i-1}, y_{i-1} \right) = \frac{D}{\Sigma} \frac{\partial \Sigma}{\partial r} \frac{x_{i-1}}{r_{i-1}}
\end{equation}
where $D$, $\Sigma$, and the derivative of the surface density are evaluated at the point ($x_{i-1},y_{i-1}$); and a contribution due to the gradient in the diffusivity:
\begin{equation}
v_{FP,x}\left( x_{i-1}, y_{i-1} \right) =\frac{\partial D}{\partial r} \frac{x_{i-1}}{r_{i-1}}
\end{equation}
The dependency in the $y$ direction would be similar as above.  Note that we are ignoring orbital motions here, as we are assuming axial symmetry--the only concern is variations with distance from the origin.  Were we concerned with azimuthal variations, orbital velocities would have to be added to the expressions given above.

To validate this model, we compare the results of this particle-tracking model to those that solve the radial advection-diffusion equation Eq. (4) using a finite-volume method.  The so called ``dye-tracking" approach, which tracks how the abundance of a species changes as a function of location and time, has been the primary method by which transport has been investigated in previous studies \citep[e.g.][among many others]{gail01,bockelee02, cuzzi03, kg04,boss08, ciesla07,ciesla09,ciesla10cai}.  The dye-tracking model calculations in this study were performed using the model described in \citet{ciesla10cai}.

The first validation test  compared  the predictions of the two models for the case of transport in a steady-state protoplanetary disk around a solar-mass star, whose physical structure was given by:
\begin{equation}
\Sigma \left( r \right) =  2000 \left( \frac{r}{\mathrm{1 AU}}\right)^{-1} ~\mathrm{g/cm}^{2}
\end{equation}
\begin{equation}
T \left( r \right) = 280 \left( \frac{r}{\mathrm{1 AU}}\right)^{-\frac{1}{2}} ~\mathrm{K}
\end{equation}
As we have assumed a standard $\alpha$-disk in steady-state, the advective velocity everywhere was given by $v_{r}$=-3$\nu$/2$r$, where $\nu$ is the local viscosity.  A value of $\alpha$=10$^{-3}$ was assumed, and all particles were taken to be 1$\mu$m in radius.

We tracked the transport of the trace species in the disk using both types of transport models described above.  The results are shown in Figure 1.  The trace material in the dye-tracking model was released in the grid  that spanned $\sim$5.0 to $\sim$5.4 AU (centered at 5.2 AU), such that $\Sigma_{i}$=1 g/cm$^{2}$ at the beginning of the simulation.   We then calculated how $\Sigma_{i}$ evolved with time and location.  The results for this model at $t$=10$^{4}$, 10$^{5}$, and 10$^{6}$ years are shown as the solid lines.

In the case of the particle-tracking model, $N$=10,000 particles were released at 5.2 AU at the beginning of the simulation.  Each particle thus represented $A \Sigma_{i}$/$N$=9.36$\times$10$^{22}$ g of trace materials, where $A$ represents the area of the annulus spanning from $\sim$5.0 to $\sim$5.4 AU.  In calculating the evolution of the trace particles, Equations (6) and (7) were used, with the local values of the gas determining the appropriate velocities and diffusive step.  The timestep for each particle was assumed to be $\delta t$=0.1/$\Omega_{K}$, where $\Omega_{K}$ is the local orbital angular velocity at the location of the particle at the beginning of the time step.   This is roughly the correlation time expected for nebular turbulence \citep{fromang06}.  Particles were released at $t$=0 and tracked for a period of 10$^{6}$ years, with their locations recorded at 10$^{4}$ and 10$^{5}$ years.  When the timestep would have pushed the total simulation time for a given particle beyond the time of interest, the timestep was was reduced to allow the locations to be recorded.  In calculating the surface density of the trace species, the number of particles located between different grid points used in the dye-tracking calculations were determined, and the total mass represented by these particles then divided by the effective area of the annulus in which they were located.  These surface densities distributions are shown as dashed lines in Figure 1.  

We also considered transport in a viscously evolving disk, thus relaxing the assum:ption of a steady-state disk. 
For this case we started with a 0.1$M_{\odot}$ disk with an initial surface density of
\begin{equation}
\Sigma \left( r \right) =  14250 \left( \frac{r}{\mathrm{1 AU}}\right)^{-1} ~\mathrm{g/cm}^{2}
\end{equation}
out to 10 AU. Again, the standard $\alpha$-disk model was used, with the model of \citet{ciesla10cai} used to calculate the temporal evolution of the disk along with the transport of materials within it.  Trace materials were tracked in the same manner and with the same initial conditions as described in the steady-state case.  These results are shown as the solid lines in Figure 2.

The particle-tracking calculations were performed as described above, however, the changing dynamic environment within the disk had to be accounted for.  The properties of the disk (surface density, temperature, etc.) were recorded at every grid point in the disk model at 100 year intervals in the case of the evolving disk.  At each timestep in the particle tracking model, the properties of the gas and dynamics of the disk at the location of interest were found by linear interpolation from this spatial-temporal map of disk evolution. The results of the particle-tracking model in the evolving disk are again shown as the dashed lines in Figure 2.

As can be seen, there is strong agreement between the two  different transport models  in cases considered here.  One immediately notes that the inferred surface densities for the particle-tracking model at 10$^{6}$ years are not as smooth as the earlier times within each case.  This is  due to trace materials (particles) being lost from the disk over time as they are accreted by the central star.  This results in a smaller number of particles surviving at these later stages.  These smaller number of surviving particles are thus not sufficient to provide a detailed description of the collective behavior of all particles in the disk, though the remaining particles still reproduce the general distribution quite well.  This agreement demonstrates that the particle-tracking model provides a good representation of the motions of particles that are subjected to the diffusive and advective motions expected in the model disks described here.

\section{Particle Paths and Environments Seen}

The advantage of the particle-tracking model over traditional dye-tracking models is that it allows one to investigate the particular paths that materials would take over their lifetimes in a protoplanetary disk.  This is demonstrated in
Figure 3, which shows the paths of 3 particles from the simulation of the evolving disk described above.  These particles originated at the same location in the disk (5.2 AU), and each finished the simulation (after 10$^{6}$ years)  located within $<$0.01 AU of one another at $\sim$2.56 AU.  As can be seen, however, each particle took very different paths to reach that final location, with each moving as far as 13, 22, and 27 AU from the central star over their lifetimes.   Given that the particles migrate through different regions of the disk during their 1 million year disk lifetime, they will see different thermal, pressure, and irradiation environments from one another. This would allow each grain to be processed differently within the disk prior to meeting up again, where they could potentially accrete into a common parent body.

Figure 4 shows the different paths of the particles during the first 10$^{5}$ years of the simulation, both in terms of their distances from the star, as well as the ambient temperatures that they would see.  While two particles are pushed outwards fairly rapidly, moving from $\sim$5 to 15 and 20 AU in this 10$^{5}$ year period, the third particle, whose path is indicated by the dark, black line in Figure 4, moves around the inner region (1-8 AU) of the disk over this same time period.  As such, the two outwardly drifting particles move to cooler regions of the disk and generally never see temperatures above $\sim$150 K after 40,000 years of model time.  The third particle, however, spends the first 10$^{5}$ years at temperatures above 200 K, with a number of excursions into regions of the disk with temperatures above 600 K, including seeing temperatures above 1000 K multiple times in the first 10$^{4}$ years.  Such temperatures are high enough that grains could have been subjected to physical processes such as annealing, or undergone chemical alteration through rapid reactions with the gas--processes that the other two grains would have escaped because they never saw temperatures this high.

While these concepts are not necessarily new, the paths of the three particles shown in Figures 3 and 4 illustrate the importance of understanding what paths particles took over their lifetimes in a protoplanetary disk--it is not enough to assign the chemical compositions or physical properties of materials based on their initial or final locations in the disk.  Figure 5 shows the range of locations seen by the 3803 surviving (out of 10$^{4}$ modeled) particles that reside outside of 1 AU after 10$^{6}$ years of evolution in the protoplanetary disk (and did not migrate within 0.1 AU of the star at any point).  Plotted are the maximum and minimum radial distances from the star for each particle as a function of their final location in the disk for both the entire disk (left panel) and for those that finish the simulation within 5 AU of the star (right panel).  For any given location, the materials present at the end of the simulation have seen a range of distances from the star, and thus, physical environments within the disk.  For example, in the asteroid belt region, roughly the 2-4 AU region of the disk, we see some grains have moved inwards as far as 0.2 AU in the disk, while some have never gotten closer than their final location of 2-4 AU.  These same grains typically migrated outwards beyond 10 AU, with some seeing distances as far out as 40 AU before migrating inwards again.

The various distances seen by the particles at a given location would suggest different thermal histories for those grains.  Figure 6 shows the highest temperature each particle saw in the disk plotted as a function of its final radial location.  Those particles that remain in the inner disk ($<$ 5 AU) have seen a wide range of peak temperatures, from $\sim$620 K, corresponding to the temperature of the disk at their point of origin at the start of the simulation, to $\sim$ 2000 K, which is achieved much closer to the star (most grains would have been vaporized at such temperatures, but we ignore this effect for now--we return to this issue below).  This same range of temperatures is seen in those particles that finish the simulation inside of 50 AU, though there is a trend that the further out in the disk that a particle is located at the end of the simulation, the lower the range of peak temperatures seen.  Specifically, at greater distances from the central star, the lower the maximum temperature seen by the particles present there.  This is due to the fact that the particles that migrate furthest outwards in the disk do so by having short residence times in the inner disk, and thus are able to ride the viscous expansion of the disk (residing in the region of the disk where the net motions are outward) to very large distances from the star. As a result, grains that start at a given location and then wind up in the inner disk will have been exposed to a wider range of peak temperatures, whereas those grains in the outer disk rarely saw temperatures in excess of the temperatures present in the disk where they originated.

It is important to remember that the results shown here consider only the dynamics of those dust particles released at a given location in the disk.  This point is selected to reside far enough out in the disk that temperatures are too low for any significant annealing to take place and thus the grains are assumed to begin amorphous, which is important for the discussion below.  While the specifics of the dynamical evolution of particles released at other locations are expected to differ from that here, the general trends should remain the same.

Figures 7 and 8 show the radial ranges  and peak temperatures seen of particles as Figures 5 and 6, but for a disk in which the turbulent parameter, $\alpha$, was set to 10$^{-4}$ (with the same initial mass distribution).   Particles were released at the same location in the disk as the case just discussed, 5.2 AU, which is at a lower temperature than above, $\sim$250 K due to the lower levels of viscous dissipation.  Because the level of turbulence is lower, the random motions of the particles  in a given time interval are smaller than the preceding case.  Thus the range of behavior for particles that wind up at a given location in the disk is smaller than in the previous case.  That is, those particles that finish the simulation between 2 and 4 AU rarely migrated outwards of 10 AU from the central star, while still exhibiting a similar range of minimum radii as above.  

These results illustrate that each particle follows its own, independent path within a diffusive protoplanetary disks,  and thus sees a unique integrated set of protoplanetary disk environments.  As such, those particles that are found in close proximity to one another at a given time would have seen very different conditions within the disk, 
which would translate to different chemical and physical evolutionary histories. The particular ranges of conditions  seen by a collection of particles will depend on the level of turbulence (diffusion) in the disk.   The differences in disk conditions seen by ``sibling'' particles would not be revealed in the dye-tracking models traditionally used in transport studies, and can only be quantified by following the representative paths that particles take through the disk as done here.

\section{Application to Annealing of Amorphous Dust}

Outward transport from the hot, inner regions of the solar nebula has  been a focus of recent work as it may offer a way of explaining the presence of crystalline silicates in comets and the outer regions of protoplanetary disks \citep{nuth00,nuth06,wooden08,bockelee02,gail01,ciesla07,ciesla09}.  That is, silicates that were present in the parent cloud core from which our solar system formed are expected to have been predominately amorphous, based on the high abundance of such grains in the diffuse ISM \citep{kemper04}.  Such grains would remain amorphous once incorporated into the solar nebula, unless exposed to high enough temperatures that they could undergo annealing \citep{fabian00,hallenbeck00,djouadi05,murata07,roskosz09,roskosz11}.  The level to which such grains were transformed depends on the temperatures that the solid particles were exposed to within the solar nebula and how long they spent at the different temperatures.  

 Previous studies for the formation and redistribution of crystalline grains in the solar nebula have generally accounted for annealing by assuming precursors become crystalline upon entering a region of the disk that was above some critical temperature, $T_{a}$.  This temperature is generally defined as the temperature where the annealing timescale was 
sufficiently small that it could be considered instantaneous \citep[e.g.][]{gail01,bockelee02,dullemond06,ciesla07}.  However, annealing could be a drawn out process, occurring over an extended period at lower temperatures.
  Further, it is unclear what the extent of the annealing would be for each grain in these studies--that is, how often are there grains that undergo partial annealing, but do not become fully crystalline?  These issues can only be addressed by tracking the exact temperature-time history of grains in a protoplanetary disk and applying the models for annealing rates that have been derived from laboratory studies.

While  annealing is recognized to require high temperatures, the exact temperatures and rates at which this transformation occurs depend on the structure and composition of the amorphous precursor.  Experimental studies on Mg-rich olivine and chondritic minerals \citep{fabian00,djouadi05,murata07} have shown that the annealing process is well described by the Johnson-Mehl-Avrami (JMA) equation \citep{burke65}:
\begin{equation}
X= 1-\mathrm{exp}\left[ -\left(\frac{t}{\tau}\right)^{n}\right]
\end{equation}
where $X$ is the fraction of displaced atoms that are transformed (annealed) in a given time, $t$, and $n$ is an exponent that depends on the crystallization kinetics.  The characteristic time constant of transformation, $\tau$, is dependent on the temperature and follows an Arrhenius relation:
\begin{equation}
\tau^{-1} = \nu e^{-\frac{E_{a}}{kT}}
\end{equation}
where $\nu$ is a characteristic vibrational frequency and $E_{a}$ is the activation energy.  While experimental results vary somewhat, typical values of the parameters for amorphous Mg$_{2}$SiO$_{4}$ smokes are are $n$=1, $\nu$=2$\times$10$^{13}$ s$^{-1}$, and $E_{a}$/$k$=39,100 K \citep{fabian00}, with values of $\nu$=2.5x10$^{26}$ s$^{-1}$ and $E_{a}$/$k$=70,000 K estimated for MgSiO$_{3}$ glass particles (\citet{bockelee02} based on analysis of \citet{fabian00} experiments).  Importantly, \citet{djouadi05} demonstrated that the annealing of the amorphous silicates is independent of the amorphization history of the grain--that is, the annealing of the grain will proceed in the same manner regardless of whether the grain is completely amorphous from its lifetime in the parent molecular cloud or if it was cycled in and out of warm environments, partly annealing over multiple intervals throughout.

\citet{hallenbeck00} offered an alternative way of tracking the change in the structure of Mg-silicates by observing the evolution of their spectra over time as they were exposed to different temperatures over different periods of time.  Like \citet{fabian00}, they used Mg-smokes as their starting point, though  the smokes in \citet{hallenbeck00} were formed from vapor phase condensates that were quenched at low pressures, whereas \citet{fabian00} smokes were formed at slightly higher pressures.  The \citet{hallenbeck00} smokes have been referred to as ``chaotic'' as they require some warming and annealing to occur just to get to structures similar to those smokes studied in \citet{fabian00}.   \citet{hallenbeck00} noticed that the spectra of the annealing grains evolved in distinct stages, rather than continuously.  Most importantly, they described the spectral stall, where little change occurred in the 10 $\mu$m region of the spectra for grains heated to just above 1000 K, but rapid evolution was seen when slightly higher temperatures were reached.  As such, \citet{hallenbeck00} defined a two-stage annealing process that they quantified by defining the Silicate Evolution Index (SEI).  Amorphous grains would begin with SEI=0 and evolve with a given rate up until the stall point, or SEI=1.  Beyond that, the rate of evolution would change, with the spectra continuing to evolve towards that of crystalline silicates, which is quantified with the SEI increasing up to a value of 102. 

A comparison of the three annealing rates considered here is shown in Figure 9, where the time constant of transformation is plotted as a function of temperature.  While other annealing rates have been derived, they tend to be similar to or fall within the range of those shown here.  To illustrate the importance of understanding the disk environments seen by a given particle, consider the annealing rate of \citet{fabian00} (solid line).  Amorphous grains
would take $\sim$1 year to anneal at a temperature of $\sim$800 K, and such a temperature may be used as an estimate for $T_{a}$, the critical temperature for annealing, as this timescale is comparable to a typical dynamical timescale in a disk \citep[e.g. this was the criterion in ][]{dullemond06}.  However, amorphous grains brought to a temperature of 650 K would also anneal if maintained at this temperature for a period of $\sim$10$^{5}$ years.  In order to fully evaluate how annealing proceeds in a protoplanetary disk, it is necessary to quantify how long particles may reside in these lower temperature environments where annealing may still occur on timescales that are less than the typical lifetime of a protoplanetary disk \citep[of order 10$^{6}$ years;][]{haisch01}.  Note we are ignoring any mineralogical changes that may occur, though this may be important in determining the range of temperatures at which grains were annealed \citep{roskosz09,roskosz11}.

We have used the calculations in the validation cases described above to evaluate how amorphous precursors, starting at an initial temperature of $\sim$620 K (for the $\alpha$=10$^{-3}$ case), would be physically altered as a result of their motions through disk.  Each grain was assumed to be perfectly amorphous ($X_{0}$=0 or SEI$_{0}$=0) at the beginning of the simulation.   The extent of annealing for a grain was determined by integrating over its individual path in the disk, tracking the temperatures (taken as the midplane temperature) each grain saw and the time spent at each temperature.  Each of the annealing/crystallization rates identified above, the MgSiO$_{4}$ of \citet{fabian00} smokes, MgSiO$_{3}$ glass of \citet{bockelee02}, and the Mg-silicate smokes of \citet{hallenbeck00} were considered.  The results are shown in Figures 10-12, which plot the crystallinity fraction achieved by each grain as a function of the highest temperature the particle was exposed to over its nebular lifetime (left panels of each figure).   Also plotted in each figure are the crystallinity fraction (fraction of the displaced atoms that have been transformed) for each grain versus the final location of the particle at the end of the simulation (set as 10$^{6}$ years).  

For each of the annealing rates, the grain distributions form a near step function in that grains are completely amorphous (fraction annealed or SEI of zero) if peak temperatures were below a certain value, transitioning to completely crystalline (crystalline fraction of 1 or SEI of 102) over a small temperature range, of order tens of Kelvin.    This is due to the exponential dependence of the annealing laws, which cause the annealing timescales to rapidly decrease with increasing temperature. The temperatures at which the transitions from amorphous to crystalline take place are $\sim$ 700, 850, and 1000 K for the \citet{fabian00}, \citet{bockelee02}, and  \citet{hallenbeck00} annealing laws, respectively. Thus in models where kinetics are not explicitly accounted for because the paths of the particles are not tracked, such as the dye-tracking models of previous studies, these temperatures represent good approximations for the temperature at which the amorphous-to-crystalline transition is made. 

The step function behavior of annealing implies that there will be an almost binary nature of grains in the protoplanetary disk: they are generally either 0\% or 100\% crystalline, with a small fraction achieving intermediate values.  
The fraction achieving intermediate values is greatest in the \citet{fabian00} law, as displaced atoms are more mobilized at lower temperatures.   In examining grains from chondritic meteorites, interplanetary dust particles (IDPs), and cometary grains, there are very few that seem to suggest that they were only partially annealed (S. Messenger, personal communication). This suggests that the annealing laws of \citet{bockelee02} or \citet{hallenbeck00} may more accurately describe the behavior of silicates in our solar system.

This general behavior is true regardless of the size of the particles considered.  Figure 13 shows the level of annealing achieved for grains 10 $\mu$m, 100 $\mu$m, and 1 mm  in radius plotted against their final location in the disk after 10$^{6}$ years of evolution. The dynamics of the various sized particles differ primarily due to larger particles being dragged inwards due to the interactions with the gas \citep{weid77}, with the lower diffusivites of the larger particles \citep{yl07} playing a secondary role in these cases.  These effects can be seen as the 1 mm particles are all found inside of $\sim$30 AU in the disk at the end of the simulation, while the outermost distance for each other particle increases with smaller size.  These same effects also lead to a decrease in the survival frequency of the particles with increasing size.  Despite these differences, however, the fraction of grains that survive that are completely annealed is constant within the disk. Further, there still remains only a small number of particles that undergo intermediate levels of annealing, again implying grains will largely be either perfectly amorphous or perfectly crystalline.

It is worth noting again that vaporization of dust grains is not explicitly modeled here.  Olivine and pyroxene grains, the dominant silicates found in primitive bodies in our Solar System, are expected to vaporize at temperatures of $\sim$1300-1500 K within the solar nebula, with the exact value depending on the pressure of the gas \citep{ebel00,davisrichter05,grossman10}.  In reality, the grains that reached these temperatures would vaporize, and could only be preserved in the disk if the resulting vapor was carried outward again to recondense.  Indeed, a number of  those grains in the model developed here saw temperatures of $T >$1500 K, yet remain in the disk after 10$^{6}$ years, implying this would have occurred.  Upon condensation, the silicates may have been amorphous rather than crystalline, though they are treated as being crystalline in the model.  This is justified as the condensation temperature for the silicates would largely exceed 1000 K, so that even if amorphous grains condense, they would quickly anneal as shown here.  If it were only vaporization and recondensation that allowed crystalline silicates to form, the distributions of crystalline grains shown here would be similar, though with a smaller fraction of crystalline materials.  Plots of the crystallinity versus peak temperature would continue to remain as a step function, with the transition of amorphous to crystalline occurring right at the temperature where vaporization occurred (generally higher than the transition temperatures in the annealing laws here).

In addition to modeling the kinetics of the annealing process, we can also examine when the crystalline grains in the disk were annealed.  Figure 14 plots the time of crystallization of the grains, defined as the time in the model when a grain reached 100\% annelaing or SEI=102, as a function of distance from the central star.    As can be seen, silicates are annealed throughout the entire 1 million years of model time considered here --as long as there is a region in the disk that is sufficiently hot and particles are able to reach it, then annealing can occur, regardless of the annealing law used. A trend can be seen in that those grains in the outer part of the disk are annealed very early on in disk evolution, whereas those grains that are annealed at later times are found closer to the star.  This is again understood in terms of the dynamical evolution of the disk--those grains that migrate furthest outwards in the disk do so early on by taking advantage of the viscous expansion of the disk.  At later times, outward transport becomes more difficult, as the gas has a net inward flow for a larger radial expanse of the disk.  Thus any grains that are exposed to high temperatures in the inner region of the disk at later times are only able to diffuse outwards a relatively small distance before the inward flows dominate their motions.  This same effect is what \citet{ciesla10cai} argued was responsible for the narrow age distribution of CAIs in chondritic meteorites \citep{thrane06,jacobsen08}.

This implies that the crystalline grains observed in comets today should have all been annealed early in the formation of the solar system.  \citet{nuth00} proposed that the ages of comets could be inferred from the relative proportion of crystalline and amorphous silicates: ancient comets would contain largely amorphous grains, while those that formed later could incorporate more and more annealed grains brought outwards from the inner solar system.  Here we see that while grains may be delivered to the outer disk over an extended period of time, it is most likely to occur early in disk evolution.  That is, those grains that are delivered to the outer disk likely annealed during a short period of time ($<$10$^{5}$ yrs) very early in the evolution of the disk, because those grains that migrated to the high temperature region of the disk early on could anneal and be pushed out to where the comets formed by the viscous expansion of the disk.  Those grains that originated in the cool, outer regions of the disk would only anneal once they migrated to the hot, inner disk and could only be incorporated into comets if they were transported to the outer disk again.  This process becomes less and less efficient over time as discussed above--the dynamical evolution of the disk limits how much material can make it to the outer disk as it evolves.  

This behavior may result in a difference in the amount of crystalline silicates that are incorporated into Oort cloud and Kuiper Belt comets.  That is, Oort cloud comets are thought to have formed in the vicinity of Jupiter and were subsequently scattered outwards through gravitational interactions with the giant planets.  The Kuiper Belt comets, on the other hand, largely formed outside the orbits of the giant planets.  These comets thus may contain different populations of annealed grains, with those in the Oort cloud able to accrete a portion of the grains that were annealed later in disk history, whereas the Kuiper Belt comets would only be able to accrete those grains that were annealed earlier.

This result serves as a prediction of the annealing model for the source of crystalline grains in the outer disk--in the context of the model described here, all such grains should have been exposed to high temperatures very early in the evolution of the disk.  If high temperature minerals instead exhibit a wide range of ages, including some that were exposed to such temperatures a few x10$^{5}$ years into the evolution of the solar nebula, an alternative source of crystalline grains may be needed, such as the possible annealing of grains in transient heating events like  shocks \citep{harker02}.  While dating the timing of annealing of such grains may be difficult, it is worth considering the recent observations reported by \citet{oliveira11},  who compared the dust mineralogy in a number of protoplanetary disks in clusters of different ages.  It was found that all disks have roughly the same silicate crystallinity fraction, independent of age.  They used this observation to suggest that annealing of silicates would have been limited to very early in disk evolution $<$1 Myr, and then the crystallinity was ``frozen in'' for the rest of disk evolution, remaining roughly constant with time.  A similar conclusion was reached by \citet{sicilia09}, based on their study of the $\eta$ Chamaeleontis cluster.  This is consistent with the picture developed here, where the crystalline grains form early on in the high temperatures in the inner disk, and then migrate outwards to be preserved in the disk.  The crystalline fraction would not increase significantly as it would become more difficult for newly annealed grains to be preserved in the disk at later times as the dynamical environment in the disk would continuously push these grains to the inner disk.  Indeed, the results shown here suggest that the youngest crystalline grains are found only in the inner disk, allowing the outer disk to remain at a roughly uniform crystalline fraction, which is the region \citet{oliveira11} was largely probing.

\section{Discussion and Summary}

Evidence for mixing and transport is abundant in the Solar System.  Chondritic meteorites contain an array of materials that formed in different environments yet are mixed on the scale of centimeters.  These same types of materials were also found in the Comet Wild 2 samples returned by Stardust, pointing to a large-scale exchange of materials between the inner solar nebula, that is the region in which terrestrial bodies formed, and the outer nebula where icy bodies and giant planets formed.
That small solids underwent such large-scale excursions in the early Solar System implies that they would have passed through a variety of nebular environments within which the solids would have had their chemical, isotopic, and physical properties defined or altered prior to their incorporation into a meteorite parent body, comet, or planetesimal.  

Here we have provided a means for calculating representative paths that solid particles would have taken within a diffusive protoplanetary disk.  This allows us to identify the specific environments seen by a given dust particle and quantify how long it resided in each.   Such information is critical to understanding the chemical and physical evolution of a protoplanetary disk, and as such, the method developed here provides a new way of quantitatively studying the origins of primitive materials.  We find that within a turbulent, viscous protoplanetary disk, diffusion causes grains to follow a wide range of independent paths. 

We have illustrated how this method can be used by applying it to understand the annealing of amorphous silicates in a protoplanetary disk.  Three annealing rates \citep{fabian00,bockelee02,hallenbeck00} were used to explore how the dynamical evolution of individual grains impacted their physical properties.  We found that amorphous grains largely remain completely amorphous or become perfectly crystalline, and that rarely do grains fall in between.  This is true for each of the different annealing laws investigated in detail here, suggesting this is a robust result.  Those amorphous grains that become annealed generally undergo this transformation early in the evolution of the disk.  This is particularly true for those that wind up far from the star and could possibly be accreted into comets, as they require the viscous expansion of the disk to aid their outward transport.   That is, crystalline silicates in the outer disk likely formed over a relatively small window in disk history if they formed via annealing in the hot, inner solar nebula.  If their crystallization ages could be determined, this would serve as a test for this model for their formation.  If the crystallization ages spanned a wide range of time or were found to come from a  late period in the history of the disk, this would suggest transient heating events were likely a factor in their origin.  We have considered a number of other disk structures, evolutionary parameters, and starting locations not shown here and find that these results are robust.  This may explain the reported invariance of silicate crystallinity with disk age as reported by \citet{oliveira11}.

The disk model used here used the standard $\alpha$-viscosity prescription to describe the physical evolution of the disk as it underwent mass and angular momentum transport.  The disk model thus underwent relatively ``smooth'' evolution, with the mass accretion rate of material from the disk to the central star monotonically decreasing with time.  Observations of protoplanetary disks suggest that accretion, particularly early on, was more episodic, with the rate at which material falls onto the star varying by orders of magnitude on timescales short compared to the lifetime of the protoplanetary disk \citep{hartmann96}.  Such outbursts have recently been modeled in the context of an $\alpha$-viscosity disk, where $\alpha$ varied with location and time, as its value was dependent on the local properties of the disk \citep{zhu10}. The outbursts predicted in the \citet{zhu10} study would have been most numerous during the very early stages (first few $\sim$10$^{5}$ years of evolution), and decayed in intensity and frequency with time.  The outbursts would have led to larger temperatures within the disk, allowing greater volumes of dust to be annealed than when more ``normal'' disk evolution was occurring.  As outbursts were largely limited to early times, this would again suggest that the greatest amount of dust annealing would occur during this early epoch of disk evolution. The efficiency at which such materials are transported outward and preserved in the disk, however remains to be evaluated. The differences between the $\alpha$=10$^{-3}$ and $\alpha$=10$^{-4}$ runs in this study suggest that a modest decrease in the rate of disk evolution reduces the outward mobility of small dust particles.  As such, the amount of crystalline silicates found in the cold, outer regions likely depends on the number of outbursts a given protoplanetary disk experienced, with larger numbers allowing greater amounts of dust to be processed and for greater levels of outward transport to occur.  This, however, is somewhat speculative and should be the focus of future work.  

Another issue to explore in future studies is the vertical motions of the solid materials during transport and how they would affect the inferred mineralogy of disks, since observationally, we are only able to see the very surfaces of a protoplanetary disk.  An issue to consider is whether the dust at the disk surface layer would be representative of the bulk mineralogy of the disk interior, where planet formation is expected to occur.  Vertical diffusion would carry processed grains up to the surface of the disk where they could be observed.    \citet{ciesla10} showed that at a given location in the disk, for the case of a vertically uniform $\alpha$, the dust in the upper layers of the became representative of the entire column of dust on a timescale of $t \sim H^{2}$/$D_{z}$ where $H$ was the local disk scale height and $D_{z}$ the local vertical diffusion coefficient.  For standard simplifications of $D_{z}$=$\alpha c_{s} H$, this gives $t \sim$1/$\alpha \Omega$ where $\Omega$ is the local orbital frequency.  The story becomes more complicated, however, if the vertical diffusivity varies with height above the disk midplane, as would be expected in disks where the magneto-rotational instability (MRI) is either active throughout \citep[e.g.][]{fromang06} or if it is limited to just the surface layer over an otherwise ``Dead Zone'' \citep[e.g.][]{gammie96,turner10,zhu10}.  In these cases, \citet{ciesla10} argued that the grains in upper layers would become ``trapped'' there by the turbulent eddies for long periods of time, spending extended periods of time in these high altitudes before returning back to the deeper regions of the disk.  As such, there would be less direct communication between the disk midplane and the surface layers, offering the possibility of the chemical and mineralogical species seen at high altitudes differing from that what was present in the larger fraction of the disk.  Further, here we used the simple 1D $\alpha$-viscosity disk model in order to illustrate how the particle-tracking techniques could be applied in an evolving disk.  This model assumes no significant variation in the radial transport of materials with height above the disk midplane.  Previous studies had shown that in a purely viscous disk, radial transport varied significantly with height above the disk midplane \citep{urpin84,tl02,kg04,ciesla07,ciesla09}.  In these models, outward movement is more efficient at the disk midplane than the surface layers, and \citet{ciesla07} and \citet{ciesla09} explicitly showed how this could lead to differences in the properties of materials at the disk surface versus that in the interior.  While \citet{fromang11} has found such ordered flows do not necessarily develop in fully MRI driven disks, the possibility of variations in radial transport with height do remain.  This issue requires a detailed understanding of the large-scale flows that develop in the disk, and their variations with time, and should be the focus of future work.

The results of this work also have important implications for the use of the meteoritic record to infer the types of processing that occurred in our solar nebula.  Given the wide variety of paths that solids may take in the solar nebula, and thus the range of conditions that a swarm of ``sibling'' particles (those that exist at the same location in the disk at a given time) may have seen, it is necessary to collect data on a large number of particles in order to say anything conclusive about the origin of the group as a whole with any confidence.  We must identify what properties are shared among a majority of primitive objects in the meteorites and which are limited to a small fraction.  Those rare or extreme properties may be attributed to those particles that take uncommon paths through the nebula.  Thus we must take caution in inferring properties of the solar nebula or the origin of primitive materials that are based on observations of a single or few particles.

The author is grateful for detailed comments and suggestions from Joe Nuth and an anonymous referee that led to substantial improvements in this paper.  This work was supported by NASA Grant NNX08AY47G awarded to F.J.C.

\bibliographystyle{apj}

\begin{thebibliography}{0}
\expandafter\ifx\csname natexlab\endcsname\relax\def\natexlab#1{#1}\fi

\end{thebibliography}


\begin{thebibliography}{56}
\expandafter\ifx\csname natexlab\endcsname\relax\def\natexlab#1{#1}\fi

\bibitem[{Apai {et~al.}(2005)Apai, Pascucci, Bouwman, Natta, Henning, \&
  Dullemond}]{apai05}
Apai, D., Pascucci, I., Bouwman, J., Natta, A., Henning, T., \& Dullemond,
  C.~P. 2005, Science, 310, 834

\bibitem[{Bockel\'{e}e-Morvan {et~al.}(2002)Bockel\'{e}e-Morvan, Gautier,
  Hersant, HurŽ, \& Robert}]{bockelee02}
Bockel\'{e}e-Morvan, D., Gautier, D., Hersant, F., HurŽ, J.~M., \& Robert, F.
  2002, Astronomy and Astrophysics, 384, 1107

\bibitem[{Boss(2008)}]{boss08}
Boss, A.~P. 2008, Earth and Planetary Science Letters, 268, 102

\bibitem[{Brownlee {et~al.}(2006)Brownlee, Tsou, AlŽon, Alexander, Araki,
  Bajt, Baratta, Bastien, Bland, Bleuet, Borg, Bradley, Brearley, Brenker,
  Brennan, Bridges, Browning, Brucato, Bullock, Burchell, Busemann,
  Butterworth, Chaussidon, Cheuvront, Chi, Cintala, Clark, Clemett, Cody,
  Colangeli, Cooper, Cordier, Daghlian, Dai, D'Hendecourt, Djouadi, Dominguez,
  Duxbury, Dworkin, Ebel, Economou, Fakra, Fairey, Fallon, Ferrini, Ferroir,
  Fleckenstein, Floss, Flynn, Franchi, Fries, Gainsforth, Gallien, Genge,
  Gilles, Gillet, Gilmour, Glavin, Gounelle, Grady, Graham, Grant, Green,
  Grossemy, Grossman, Grossman, Guan, Hagiya, Harvey, Heck, Herzog, Hoppe,
  Hšrz, Huth, Hutcheon, Ignatyev, Ishii, Ito, Jacob, Jacobsen, Jacobsen,
  Jones, Joswiak, Jurewicz, Kearsley, Keller, Khodja, Kilcoyne, Kissel, Krot,
  Langenhorst, Lanzirotti, Le, Leshin, Leitner, Lemelle, Leroux, Liu, Luening,
  Lyon, MacPherson, Marcus, Marhas, Marty, Matrajt, McKeegan, Meibom, Mennella,
  Messenger, Messenger, Mikouchi, Mostefaoui, Nakamura, Nakano, Newville,
  Nittler, Ohnishi, Ohsumi, Okudaira, Papanastassiou, Palma, Palumbo, Pepin,
  Perkins, Perronnet, Pianetta, Rao, Rietmeijer, Robert, Rost, Rotundi, Ryan,
  Sandford, Schwandt, See, Schlutter, Sheffield-Parker, Simionovici, Simon,
  Sitnitsky, Snead, Spencer, Stadermann, Steele, Stephan, Stroud, Susini,
  Sutton, Suzuki, Taheri, Taylor, Teslich, Tomeoka, Tomioka, Toppani,
  Trigo-Rodr'guez, Troadec, Tsuchiyama, Tuzzolino, Tyliszczak, Uesugi, Velbel,
  Vellenga, Vicenzi, Vincze, Warren, Weber, Weisberg, Westphal, Wirick, Wooden,
  Wopenka, Wozniakiewicz, Wright, Yabuta, Yano, Young, Zare, Zega, Ziegler,
  Zimmerman, Zinner, \& Zolensky}]{brownlee06}
Brownlee, D., Tsou, P., AlŽon, J., Alexander, C. M. O.~D., Araki, T., Bajt,
  S., Baratta, G.~A., Bastien, R., Bland, P., Bleuet, P., Borg, J., Bradley,
  J.~P., Brearley, A., Brenker, F., Brennan, S., Bridges, J.~C., Browning,
  N.~D., Brucato, J.~R., Bullock, E., Burchell, M.~J., Busemann, H.,
  Butterworth, A., Chaussidon, M., Cheuvront, A., Chi, M., Cintala, M.~J.,
  Clark, B.~C., Clemett, S.~J., Cody, G., Colangeli, L., Cooper, G., Cordier,
  P., Daghlian, C., Dai, Z., D'Hendecourt, L., Djouadi, Z., Dominguez, G.,
  Duxbury, T., Dworkin, J.~P., Ebel, D.~S., Economou, T.~E., Fakra, S., Fairey,
  S. A.~J., Fallon, S., Ferrini, G., Ferroir, T., Fleckenstein, H., Floss, C.,
  Flynn, G., Franchi, I.~A., Fries, M., Gainsforth, Z., Gallien, J.~P., Genge,
  M., Gilles, M.~K., Gillet, P., Gilmour, J., Glavin, D.~P., Gounelle, M.,
  Grady, M.~M., Graham, G.~A., Grant, P.~G., Green, S.~F., Grossemy, F.,
  Grossman, L., Grossman, J.~N., Guan, Y., Hagiya, K., Harvey, R., Heck, P.,
  Herzog, G.~F., Hoppe, P., Hšrz, F., Huth, J., Hutcheon, I.~D., Ignatyev, K.,
  Ishii, H., Ito, M., Jacob, D., Jacobsen, C., Jacobsen, S., Jones, S.,
  Joswiak, D., Jurewicz, A., Kearsley, A.~T., Keller, L.~P., Khodja, H.,
  Kilcoyne, A. L.~D., Kissel, J., Krot, A., Langenhorst, F., Lanzirotti, A.,
  Le, L., Leshin, L.~A., Leitner, J., Lemelle, L., Leroux, H., Liu, M.-C.,
  Luening, K., Lyon, I., MacPherson, G., Marcus, M.~A., Marhas, K., Marty, B.,
  Matrajt, G., McKeegan, K., Meibom, A., Mennella, V., Messenger, K.,
  Messenger, S., Mikouchi, T., Mostefaoui, S., Nakamura, T., Nakano, T.,
  Newville, M., Nittler, L.~R., Ohnishi, I., Ohsumi, K., Okudaira, K.,
  Papanastassiou, D.~A., Palma, R., Palumbo, M.~E., Pepin, R.~O., Perkins, D.,
  Perronnet, M., Pianetta, P., Rao, W., Rietmeijer, F. J.~M., Robert, F., Rost,
  D., Rotundi, A., Ryan, R., Sandford, S.~A., Schwandt, C.~S., See, T.~H.,
  Schlutter, D., Sheffield-Parker, J., Simionovici, A., Simon, S., Sitnitsky,
  I., Snead, C.~J., Spencer, M.~K., Stadermann, F.~J., Steele, A., Stephan, T.,
  Stroud, R., Susini, J., Sutton, S.~R., Suzuki, Y., Taheri, M., Taylor, S.,
  Teslich, N., Tomeoka, K., Tomioka, N., Toppani, A., Trigo-Rodr'guez, J.~M.,
  Troadec, D., Tsuchiyama, A., Tuzzolino, A.~J., Tyliszczak, T., Uesugi, K.,
  Velbel, M., Vellenga, J., Vicenzi, E., Vincze, L., Warren, J., Weber, I.,
  Weisberg, M., Westphal, A.~J., Wirick, S., Wooden, D., Wopenka, B.,
  Wozniakiewicz, P., Wright, I., Yabuta, H., Yano, H., Young, E.~D., Zare,
  R.~N., Zega, T., Ziegler, K., Zimmerman, L., Zinner, E., \& Zolensky, M.
  2006, Science, 314, 1711

\bibitem[{{Burke}(1965)}]{burke65}
{Burke}, J. 1965, The Kinetics of Phase Transformation in Metals (England:
  Pergamon Press)

\bibitem[{Charnoz(2011)}]{charnoz11}
Charnoz, S., Fouchet, L., Aleon, J., \& Moreira, M. 2011, Astrophysical Journal, 737, 33.

\bibitem[{Ciesla(2007)}]{ciesla07}
Ciesla, F.~J. 2007, Science, 318, 613

\bibitem[{Ciesla(2009)}]{ciesla09}
---. 2009, Icarus, 200, 655

\bibitem[{Ciesla(2010{\natexlab{a}})}]{ciesla10cai}
---. 2010{\natexlab{a}}, Icarus, 208, 455

\bibitem[{Ciesla(2010{\natexlab{b}})}]{ciesla10}
---. 2010{\natexlab{b}}, The Astrophysical Journal, 723, 514

\bibitem[{Cuzzi {et~al.}(2003)Cuzzi, Davis, \& Dobrovolskis}]{cuzzi03}
Cuzzi, J.~N., Davis, S.~S., \& Dobrovolskis, A.~R. 2003, Icarus, 166, 385

\bibitem[{D. {et~al.}(2006)D., Jerome, John, Donald, Henner, Anna, Marc,
  Stewart, Christine, Jamie, Matthieu, Giles, Yunbin, R., Peter, D., Joachim,
  Hope, Motoo, B., Anton, A., Ming-Chang, Ian, Kuljeet, Bernard, Graciela,
  Anders, Scott, Smail, Sujoy, Keiko, Larry, Russ, O., A., Franois, Dennis,
  J., J., Rhonda, Peter, Andrew, D., Karen, Laurent, \& Zinner}]{mckeegan06}
D., M.~K., Jerome, A., John, B., Donald, B., Henner, B., Anna, B., Marc, C.,
  Stewart, F., Christine, F., Jamie, G., Matthieu, G., Giles, G., Yunbin, G.,
  R., H.~P., Peter, H., D., H.~I., Joachim, H., Hope, I., Motoo, I., B., J.~S.,
  Anton, K., A., L.~L., Ming-Chang, L., Ian, L., Kuljeet, M., Bernard, M.,
  Graciela, M., Anders, M., Scott, M., Smail, M., Sujoy, M., Keiko, N.-M.,
  Larry, N., Russ, P., O., P.~R., A., P.~D., Franois, R., Dennis, S., J.,
  S.~C., J., S.~F., Rhonda, S., Peter, T., Andrew, W., D., Y.~E., Karen, Z.,
  Laurent, Z., \& Zinner, E. 2006, Science, 314, 1724

\bibitem[{{Davis} \& {Richter}(2005)}]{davisrichter05}
{Davis}, A.~M., \& {Richter}, F.~M. 2005, {Condensation and Evaporation of
  Solar System Materials}, ed. {Davis, A.~M., Holland, H.~D., \& Turekian,
  K.~K.} (Elsevier B), 407

\bibitem[{Desch(2007)}]{desch07}
Desch, S.~J. 2007, The Astrophysical Journal, 671, 878

\bibitem[{Djouadi {et~al.}(2005)Djouadi, D'Hendecourt, Leroux, Jones, Borg,
  Deboffle, \& Chauvin}]{djouadi05}
Djouadi, Z., D'Hendecourt, L., Leroux, H., Jones, A.~P., Borg, J., Deboffle,
  D., \& Chauvin, N. 2005, Astronomy and Astrophysics, 440, 179

\bibitem[{Dubrulle {et~al.}(1995)Dubrulle, Morfill, \& Sterzik}]{dubrulle95}
Dubrulle, B., Morfill, G., \& Sterzik, M. 1995, Icarus, 114, 237

\bibitem[{Dullemond {et~al.}(2006)Dullemond, Apai, \& Walch}]{dullemond06}
Dullemond, C.~P., Apai, D., \& Walch, S. 2006, The Astrophysical Journal, 640,
  L67

\bibitem[{{Ebel} \& {Grossman}(2000)}]{ebel00}
{Ebel}, D.~S., \& {Grossman}, L. 2000, \gca, 64, 339

\bibitem[{Fabian {et~al.}(2000)Fabian, JŠger, Henning, Dorschner, \&
  Mutschke}]{fabian00}
Fabian, D., JŠger, C., Henning, T., Dorschner, J., \& Mutschke, H. 2000,
  Astronomy and Astrophysics, 364, 282

\bibitem[{{Fromang} {et~al.}(2011){Fromang}, {Lyra}, \& {Masset}}]{fromang11}
{Fromang}, S., {Lyra}, W., \& {Masset}, F. 2011, ArXiv e-prints

\bibitem[{Fromang \& Papaloizou(2006)}]{fromang06}
Fromang, S., \& Papaloizou, J. 2006, Astronomy and Astrophysics, 452, 751

\bibitem[{Gail(2001)}]{gail01}
Gail, H.~P. 2001, Astronomy and Astrophysics, 378, 192

\bibitem[{{Gammie}(1996)}]{gammie96}
{Gammie}, C.~F. 1996, \apj, 457, 355

\bibitem[{{Grossman}(2010)}]{grossman10}
{Grossman}, L. 2010, Meteoritics and Planetary Science, 45, 7

\bibitem[{{Haisch Jr.} {et~al.}(2001){Haisch Jr.}, Lada, \& Lada}]{haisch01}
{Haisch Jr.}, K.~E., Lada, E.~A., \& Lada, C.~J. 2001, The Astrophysical
  Journal, 553, L153

\bibitem[{Hallenbeck {et~al.}(2000)Hallenbeck, III, \& Nelson}]{hallenbeck00}
Hallenbeck, S.~L., III, J. A.~N., \& Nelson, R.~N. 2000, The Astrophysical
  Journal, 535, 247

\bibitem[{Harker \& Desch(2002)}]{harker02}
Harker, D.~E., \& Desch, S.~J. 2002, The Astrophysical Journal, 565, L109

\bibitem[{{Hartmann} \& {Kenyon}(1996)}]{hartmann96}
{Hartmann}, L., \& {Kenyon}, S.~J. 1996, \araa, 34, 207

\bibitem[{{Jacobsen} {et~al.}(2008){Jacobsen}, {Yin}, {Moynier}, {Amelin},
  {Krot}, {Nagashima}, {Hutcheon}, \& {Palme}}]{jacobsen08}
{Jacobsen}, B., {Yin}, Q.-Z., {Moynier}, F., {Amelin}, Y., {Krot}, A.~N.,
  {Nagashima}, K., {Hutcheon}, I.~D., \& {Palme}, H. 2008, Earth and Planetary
  Science Letters, 272, 353

\bibitem[{Jacquet {et~al.}(2011)Jacquet, Fromang, \& Gounelle}]{jacquet11}
Jacquet, E., Fromang, S., \& Gounelle, M. 2011, Astronomy and Astrophysics,
  526, L8

\bibitem[{Johansen {et~al.}(2006)Johansen, Klahr, \& Mee}]{jkm06}
Johansen, A., Klahr, H., \& Mee, A.~J. 2006, Monthly Notices of the Royal
  Astronomical Society, 370, L71

\bibitem[{Keller \& Gail(2004)}]{kg04}
Keller, C., \& Gail, H.~P. 2004, Astronomy and Astrophysics, 415, 1177

\bibitem[{Kemper {et~al.}(2004)Kemper, Vriend, \& Tielens}]{kemper04}
Kemper, F., Vriend, W.~J., \& Tielens, A. G. G.~M. 2004, The Astrophysical
  Journal, 609, 826

\bibitem[{{Kinzelbach}(1990)}]{kinzelbach90}
{Kinzelbach}, W. 1990, in {Proc. Nordic seminar on groundwater modeling},
  Randsvangen, Jevnaker, Norway, 25

\bibitem[{Mousis {et~al.}(2007)Mousis, Petit, Wurm, Krauss, Alibert, \&
  Horner}]{mousis07}
Mousis, O., Petit, J.~M., Wurm, G., Krauss, O., Alibert, Y., \& Horner, J.
  2007, Astronomy and Astrophysics, 466, L9

\bibitem[{Murata {et~al.}(2007)Murata, Chihara, Tsuchiyama, Koike, Takakura,
  Noguchi, \& Nakamura}]{murata07}
Murata, K., Chihara, H., Tsuchiyama, A., Koike, C., Takakura, T., Noguchi, T.,
  \& Nakamura, T. 2007, The Astrophysical Journal, 668, 285

\bibitem[{Nuth {et~al.}(2000)Nuth, Hill, \& Kletetschka}]{nuth00}
Nuth, J.~A., Hill, H. G.~M., \& Kletetschka, G. 2000, Nature, 406, 275

\bibitem[{Nuth \& Johnson(2006)}]{nuth06}
Nuth, J.~A., \& Johnson, N.~M. 2006, Icarus, 180, 243

\bibitem[{Oliveira {et~al.}(2011)Oliveira, Olofsson, Potoppidan, {van
  Dishoeck}, Augerau, \& Merin}]{oliveira11}
Oliveira, I., Olofsson, J., Potoppidan, K.~M., {van Dishoeck}, E.~F., Augerau,
  J.-C., \& Merin, B. 2011, The Astrophysical Journal, astro-ph, 1104.3574

\bibitem[{Pavlyuchenkov \& Dullemond(2007)}]{pavly07}
Pavlyuchenkov, Y., \& Dullemond, C.~P. 2007, Astronomy and Astrophysics, 471,
  833

\bibitem[{{Roskosz} {et~al.}(2009){Roskosz}, {Gillot}, {Capet}, {Roussel}, \&
  {Leroux}}]{roskosz09}
{Roskosz}, M., {Gillot}, J., {Capet}, F., {Roussel}, P., \& {Leroux}, H. 2009,
  \apjl, 707, L174

\bibitem[{{Roskosz} {et~al.}(2011){Roskosz}, {Gillot}, {Capet}, {Roussel}, \&
  {Leroux}}]{roskosz11}
---. 2011, \aap, 529, A111+

\bibitem[{{Sargent} {et~al.}(2009){Sargent}, {Forrest}, {Tayrien}, {McClure},
  {Watson}, {Sloan}, {Li}, {Manoj}, {Bohac}, {Furlan}, {Kim}, \&
  {Green}}]{sargent09}
{Sargent}, B.~A., {Forrest}, W.~J., {Tayrien}, C., {McClure}, M.~K., {Watson},
  D.~M., {Sloan}, G.~C., {Li}, A., {Manoj}, P., {Bohac}, C.~J., {Furlan}, E.,
  {Kim}, K.~H., \& {Green}, J.~D. 2009, \apjs, 182, 477

\bibitem[{Shu {et~al.}(1996)Shu, Shang, \& Lee}]{shu96}
Shu, F.~H., Shang, H., \& Lee, T. 1996, Science, 271, 1545

\bibitem[{{Sicilia-Aguilar} {et~al.}(2009){Sicilia-Aguilar}, {Bouwman},
  {Juh{\'a}sz}, {Henning}, {Roccatagliata}, {Lawson}, {Acke}, {Feigelson},
  {Tielens}, {Decin}, \& {Meeus}}]{sicilia09}
{Sicilia-Aguilar}, A., {Bouwman}, J., {Juh{\'a}sz}, A., {Henning}, T.,
  {Roccatagliata}, V., {Lawson}, W.~A., {Acke}, B., {Feigelson}, E.~D.,
  {Tielens}, A.~G.~G.~M., {Decin}, L., \& {Meeus}, G. 2009, \apj, 701, 1188

\bibitem[{{Takeuchi} \& {Lin}(2002)}]{tl02}
{Takeuchi}, T., \& {Lin}, D.~N.~C. 2002, \apj, 581, 1344

\bibitem[{{Thrane} {et~al.}(2006){Thrane}, {Bizzarro}, \& {Baker}}]{thrane06}
{Thrane}, K., {Bizzarro}, M., \& {Baker}, J.~A. 2006, \apjl, 646, L159

\bibitem[{{Turner} {et~al.}(2010){Turner}, {Carballido}, \& {Sano}}]{turner10}
{Turner}, N.~J., {Carballido}, A., \& {Sano}, T. 2010, \apj, 708, 188

\bibitem[{{Urpin}(1984)}]{urpin84}
{Urpin}, V.~A. 1984, \azh, 61, 84

\bibitem[{van Boekel {et~al.}(2004)van Boekel, Min, Leinert, Waters, Richichi,
  Chesneau, Dominik, Jaffe, Dutrey, Graser, Henning, de~Jong, Kšhler,
  de~Koter, Lopez, Malbet, Morel, Paresce, Perrin, Preibisch, Przygodda,
  Schšller, \& Wittkowski}]{vanboekel04}
van Boekel, R., Min, M., Leinert, C., Waters, L. B. F.~M., Richichi, A.,
  Chesneau, O., Dominik, C., Jaffe, W., Dutrey, A., Graser, U., Henning, T.,
  de~Jong, J., Kšhler, R., de~Koter, A., Lopez, B., Malbet, F., Morel, S.,
  Paresce, F., Perrin, G., Preibisch, T., Przygodda, F., Schšller, M., \&
  Wittkowski, M. 2004, Nature, 432, 479

\bibitem[{Vinkovi?(2009)}]{vinkovic09}
Vinkovi?, D. 2009, Nature, 459, 227

\bibitem[{{Watson} {et~al.}(2009){Watson}, {Leisenring}, {Furlan}, {Bohac},
  {Sargent}, {Forrest}, {Calvet}, {Hartmann}, {Nordhaus}, {Green}, {Kim},
  {Sloan}, {Chen}, {Keller}, {d'Alessio}, {Najita}, {Uchida}, \&
  {Houck}}]{watson09}
{Watson}, D.~M., {Leisenring}, J.~M., {Furlan}, E., {Bohac}, C.~J., {Sargent},
  B., {Forrest}, W.~J., {Calvet}, N., {Hartmann}, L., {Nordhaus}, J.~T.,
  {Green}, J.~D., {Kim}, K.~H., {Sloan}, G.~C., {Chen}, C.~H., {Keller}, L.~D.,
  {d'Alessio}, P., {Najita}, J., {Uchida}, K.~I., \& {Houck}, J.~R. 2009,
  \apjs, 180, 84

\bibitem[{Weidenschilling(1977)}]{weid77}
Weidenschilling, S.~J. 1977, Monthly Notices of the Royal Astronomical Society,
  180, 57

\bibitem[{{Wooden}(2008)}]{wooden08}
{Wooden}, D.~H. 2008, \ssr, 138, 75

\bibitem[{{Youdin} \& {Lithwick}(2007)}]{yl07}
{Youdin}, A.~N., \& {Lithwick}, Y. 2007, \icarus, 192, 588

\bibitem[{{Zhu} {et~al.}(2010){Zhu}, {Hartmann}, \& {Gammie}}]{zhu10}
{Zhu}, Z., {Hartmann}, L., \& {Gammie}, C. 2010, \apj, 713, 1143

\bibitem[{Zolensky {et~al.}(2006)Zolensky, Zega, Yano, Wirick, Westphal,
  Weisberg, Weber, Warren, Velbel, Tsuchiyama, Tsou, Toppani, Tomioka, Tomeoka,
  Teslich, Taheri, Susini, Stroud, Stephan, Stadermann, Snead, Simon,
  Simionovici, See, Robert, Rietmeijer, Rao, Perronnet, Papanastassiou,
  Okudaira, Ohsumi, Ohnishi, Nakamura-Messenger, Nakamura, Mostefaoui,
  Mikouchi, Meibom, Matrajt, Marcus, Leroux, Lemelle, Le, Lanzirotti,
  Langenhorst, Krot, Keller, Kearsley, Joswiak, Jacob, Ishii, Harvey, Hagiya,
  Grossman, Grossman, Graham, Gounelle, Gillet, Genge, Flynn, Ferroir, Fallon,
  Ebel, Dai, Cordier, Clark, Chi, Butterworth, Brownlee, Bridges, Brennan,
  Brearley, Bradley, Bleuet, Bland, \& Bastien}]{zolensky06}
Zolensky, M.~E., Zega, T.~J., Yano, H., Wirick, S., Westphal, A.~J., Weisberg,
  M.~K., Weber, I., Warren, J.~L., Velbel, M.~A., Tsuchiyama, A., Tsou, P.,
  Toppani, A., Tomioka, N., Tomeoka, K., Teslich, N., Taheri, M., Susini, J.,
  Stroud, R., Stephan, T., Stadermann, F.~J., Snead, C.~J., Simon, S.~B.,
  Simionovici, A., See, T.~H., Robert, F., Rietmeijer, F. J.~M., Rao, W.,
  Perronnet, M.~C., Papanastassiou, D.~A., Okudaira, K., Ohsumi, K., Ohnishi,
  I., Nakamura-Messenger, K., Nakamura, T., Mostefaoui, S., Mikouchi, T.,
  Meibom, A., Matrajt, G., Marcus, M.~A., Leroux, H., Lemelle, L., Le, L.,
  Lanzirotti, A., Langenhorst, F., Krot, A.~N., Keller, L.~P., Kearsley, A.~T.,
  Joswiak, D., Jacob, D., Ishii, H., Harvey, R., Hagiya, K., Grossman, L.,
  Grossman, J.~N., Graham, G.~A., Gounelle, M., Gillet, P., Genge, M.~J.,
  Flynn, G., Ferroir, T., Fallon, S., Ebel, D.~S., Dai, Z.~R., Cordier, P.,
  Clark, B., Chi, M., Butterworth, A.~L., Brownlee, D.~E., Bridges, J.~C.,
  Brennan, S., Brearley, A., Bradley, J.~P., Bleuet, P., Bland, P.~A., \&
  Bastien, R. 2006, Science, 314, 1735

\end{thebibliography}

\newpage
\begin{table}
\begin{tabular}{c | c | c | c}

\hline
\end{tabular}
\end{table}

\newpage
\begin{figure}
\begin{center}
\includegraphics[angle=90,width=5in]{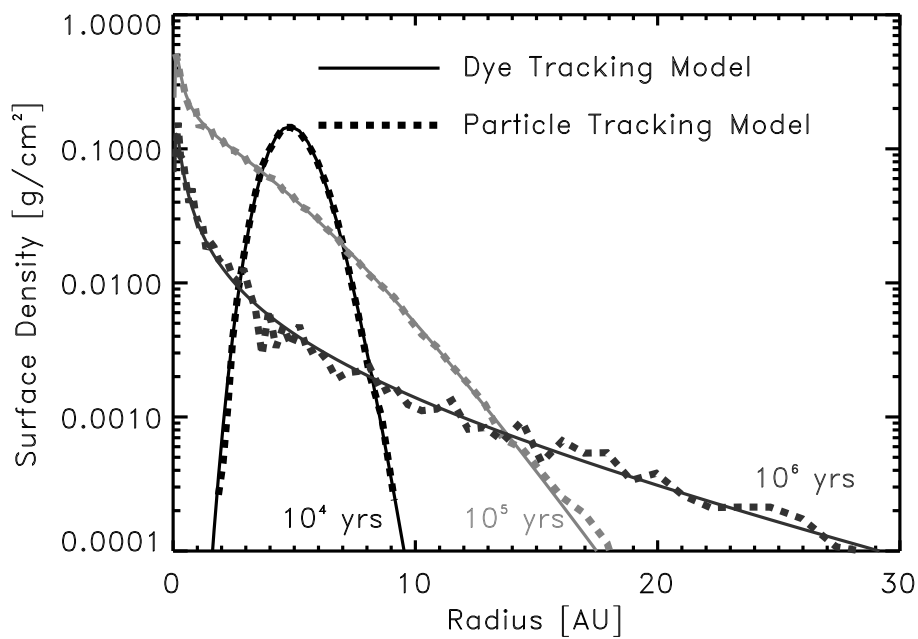}
\caption{Comparison of the particle-tracking model described here (dashed lines) with the dye-tracking model 
typically used in transport studies (solid lines).  Plotted are the surface densities of trace materials in a steady-state
disk as described in the disk, at times 10$^{4}$ (black lines), 10$^{5}$ (light grey), and 10$^{6}$ (dark grey) years.  The bumps and wiggles in the surface density distribution at 10$^{6}$ years in the particle-tracking model arises due to the fact that many particles are lost from the disk with time, leaving a small number behind to define the distribution. }
\end{center}
\end{figure}

\newpage
\begin{figure}
\begin{center}
\includegraphics[angle=90,width=5in]{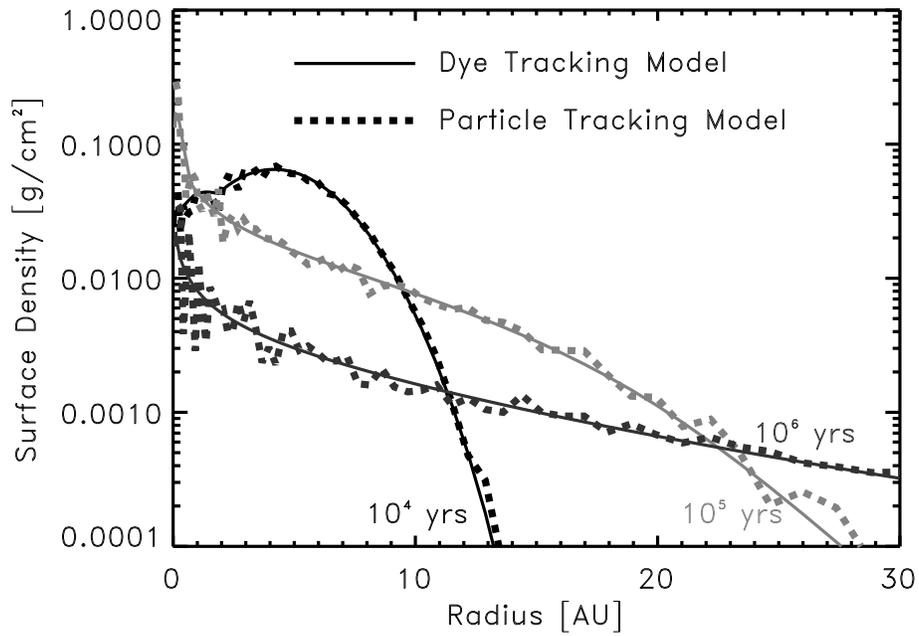}
\caption{As in Figure 1, except for the case of the evolving disk as described in the disk.  Here, just like in Figure 1, the good agreement between the two models demonstrates the particle-tracking model accurately describes the collective dynamical behavior of the particles in a protoplanetary disk. }
\end{center}
\end{figure}

\newpage
\begin{figure}
\begin{center}
\includegraphics[angle=90,width=5in]{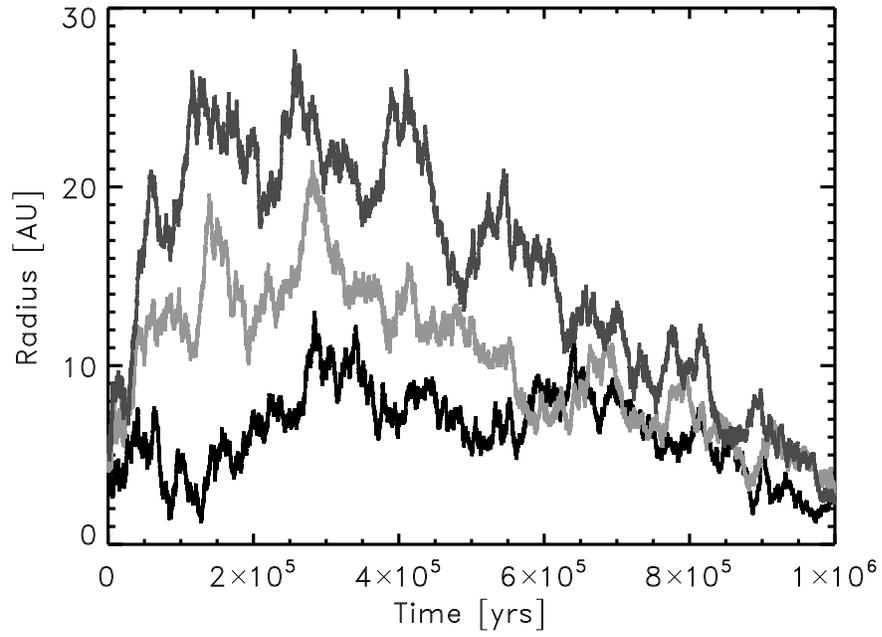}
\caption{The radial locations of 3 particles from the simulation from the evolving disk described above.  These particles originate at the same location in the disk and finish the simulation (at $t$=10$^{6}$ years) at roughly the same location in the disk $\sim$2.56 AU.  As can be seen, while their point of origin and final location are similar, each particle takes a different path through the disk, which would lead to exposure to and residence in different protoplanetary disk environments.}
\end{center}
\end{figure}

\newpage
\begin{figure}
\begin{center}
\includegraphics[angle=90,width=3.1in]{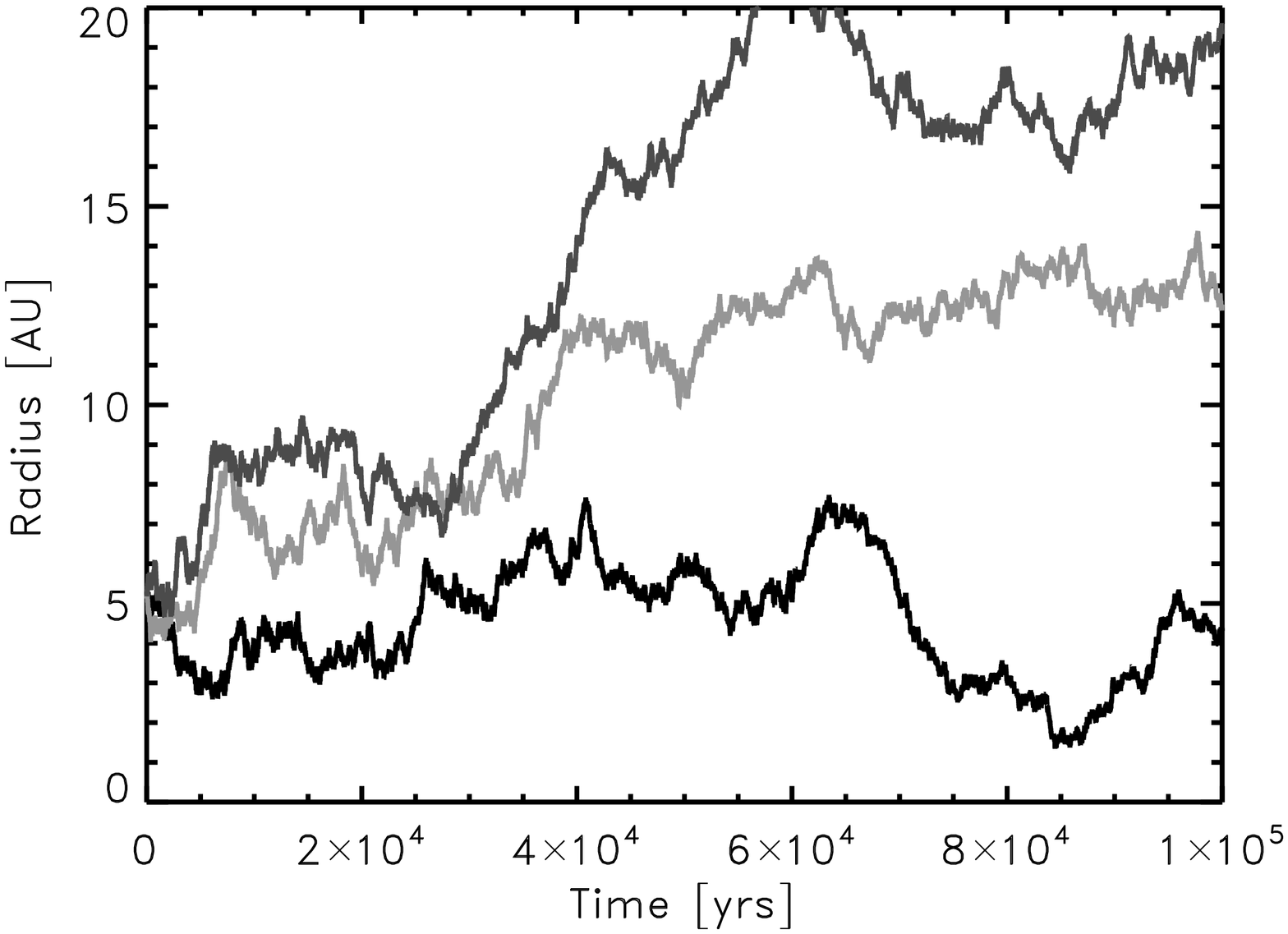}
\includegraphics[angle=90,width=3.1in]{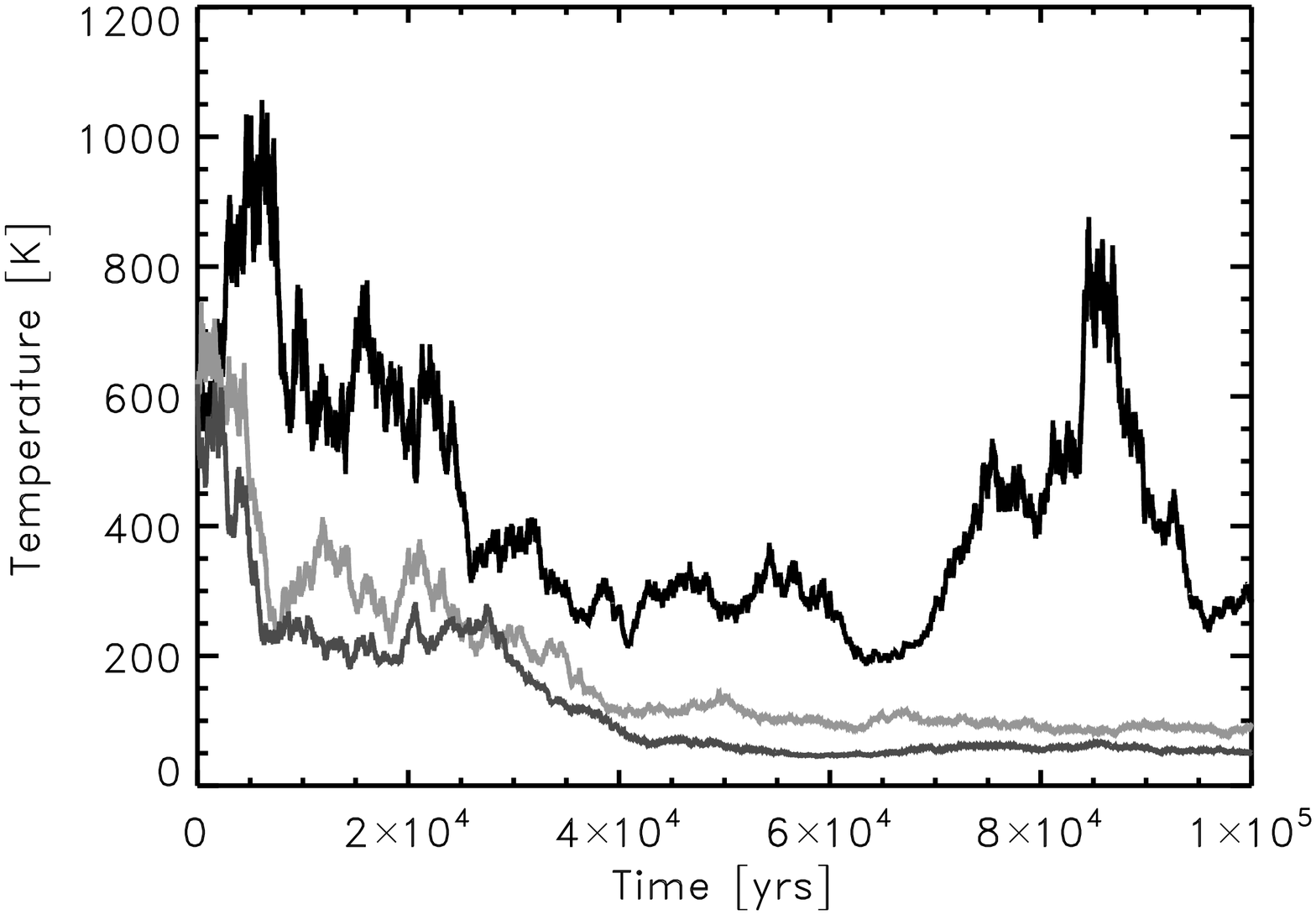}
\caption{\emph{Left:} The paths of the particles shown in Figure 3 during the first 10$^{5}$ years.  \emph{Right:}The corresponding temperatures seen by these particles in that time.}
\end{center}
\end{figure}

\newpage
\begin{figure}
\begin{center}
\includegraphics[angle=90,width=3.1in]{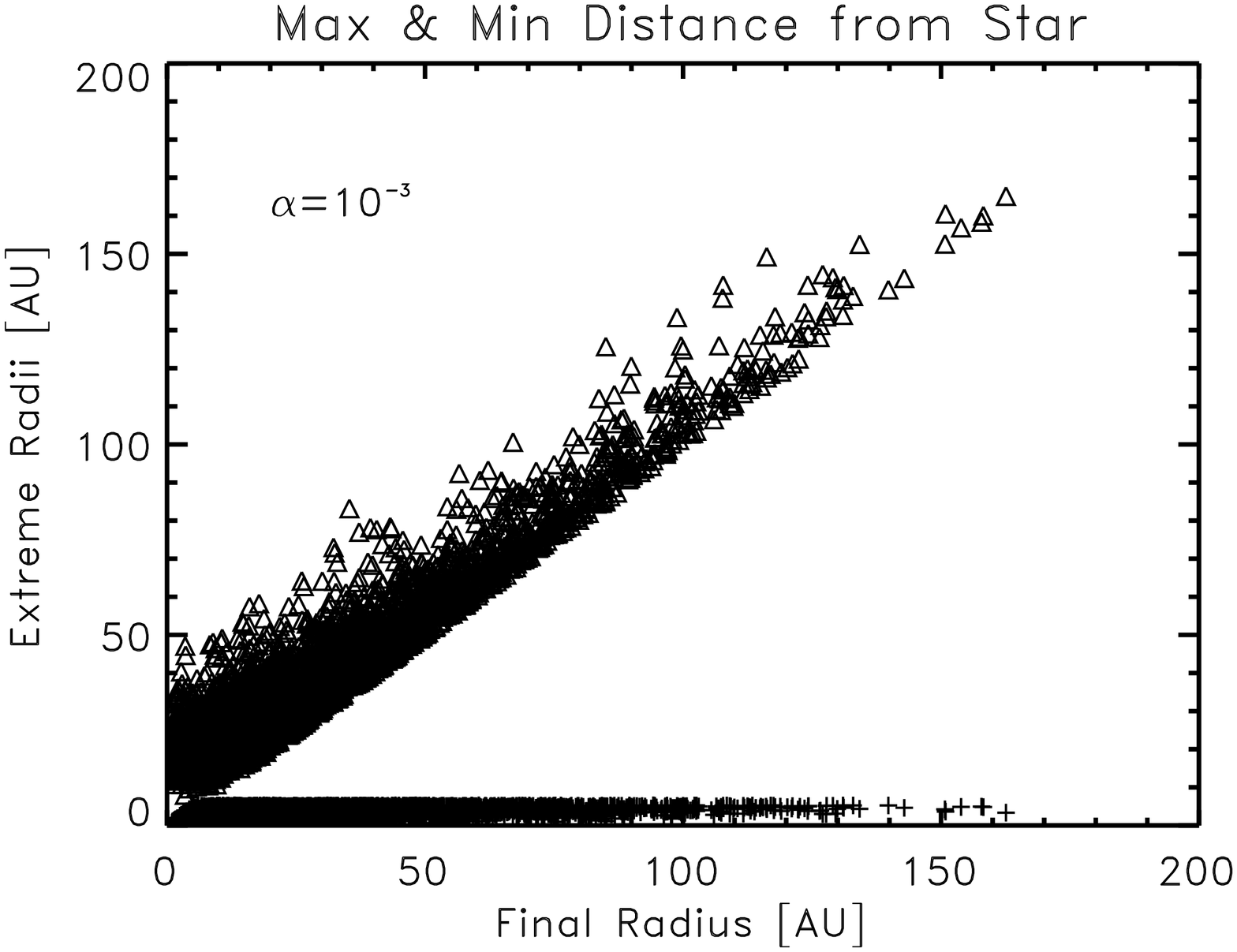}
\includegraphics[angle=90,width=3.1in]{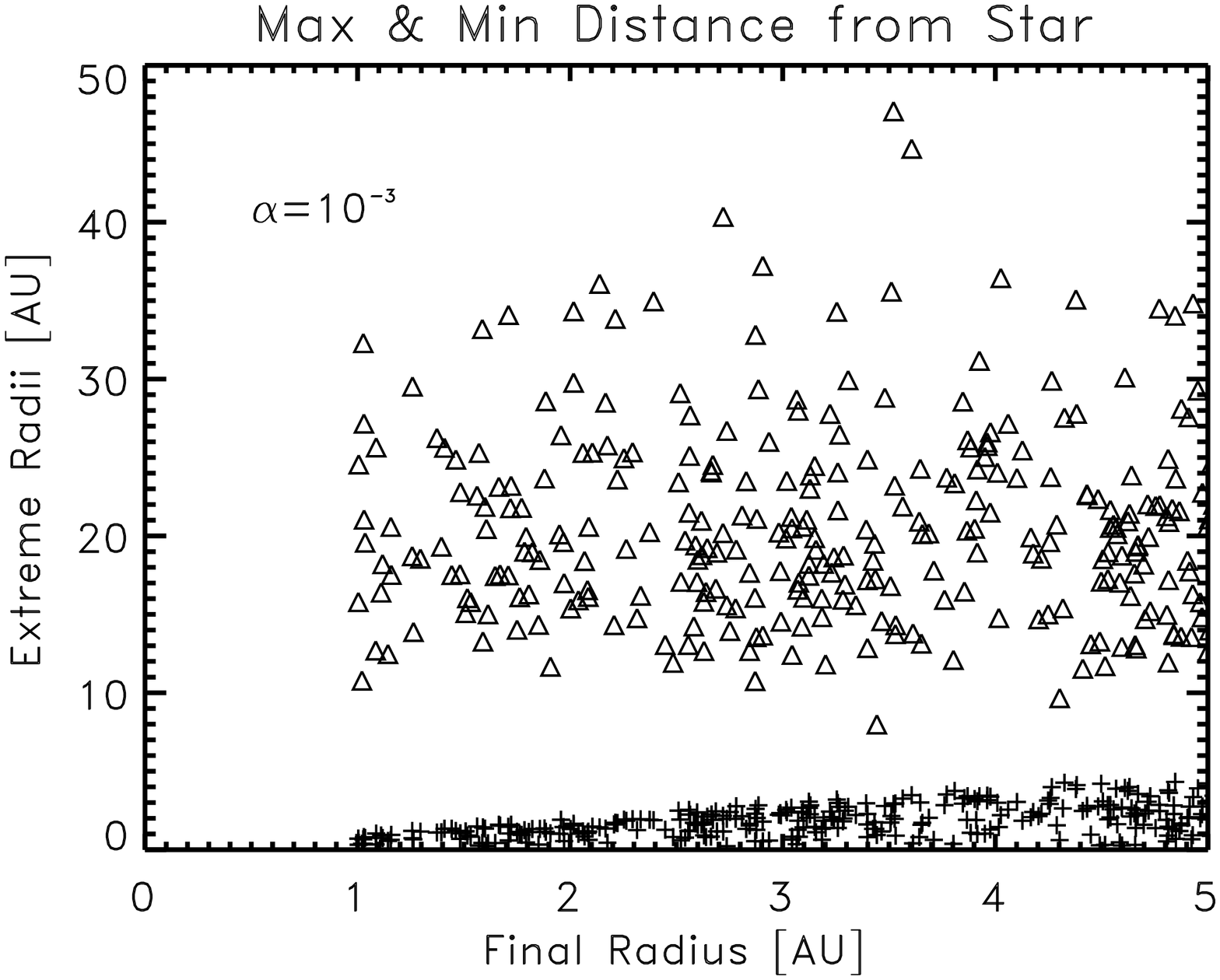}
\caption{Plotted are the extreme radii (outermost=triangles and innermost=crosses) for each of the surviving particles in the evolving disk simulation described above plotted as a function of the final position of the particles at the end of the simulation.  \emph{Left:} All particles remaining in the disk. \emph{Right:} Those particles whose final position ranges from 1 to 5 AU.}
\end{center}
\end{figure}

\newpage
\begin{figure}
\begin{center}
\includegraphics[angle=90,width=3.1in]{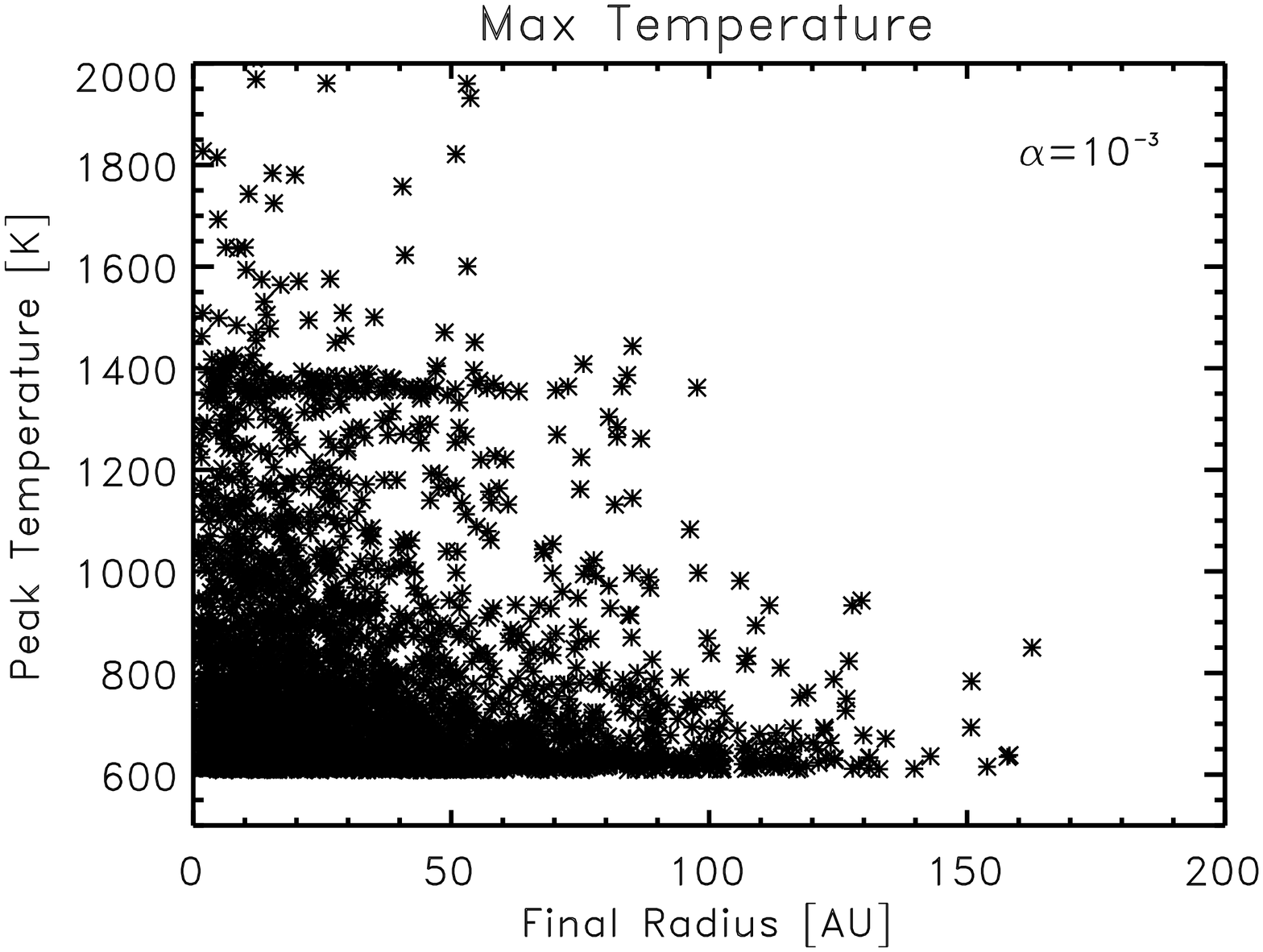}
\includegraphics[angle=90,width=3.1in]{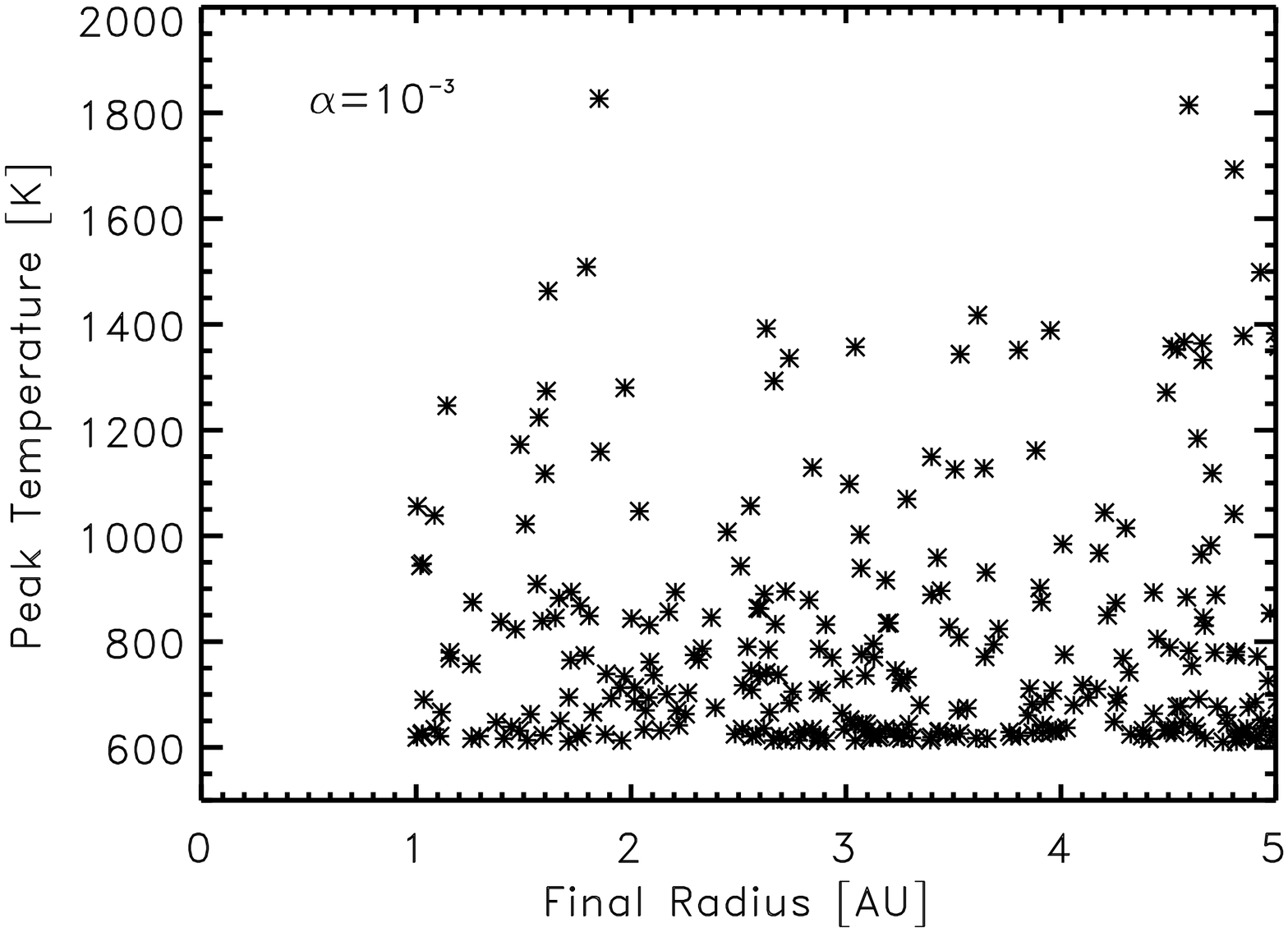}
\caption{Maximum temperature seen by each surviving particle plotted versus the final position of those particles at the end of the simulation.  \emph{Left:} All particles remaining in the disk. \emph{Right:} Those particles whose final position ranges from 1 to 5 AU.}
\end{center}
\end{figure}

\newpage
\begin{figure}
\begin{center}
\includegraphics[angle=90,width=3.1in]{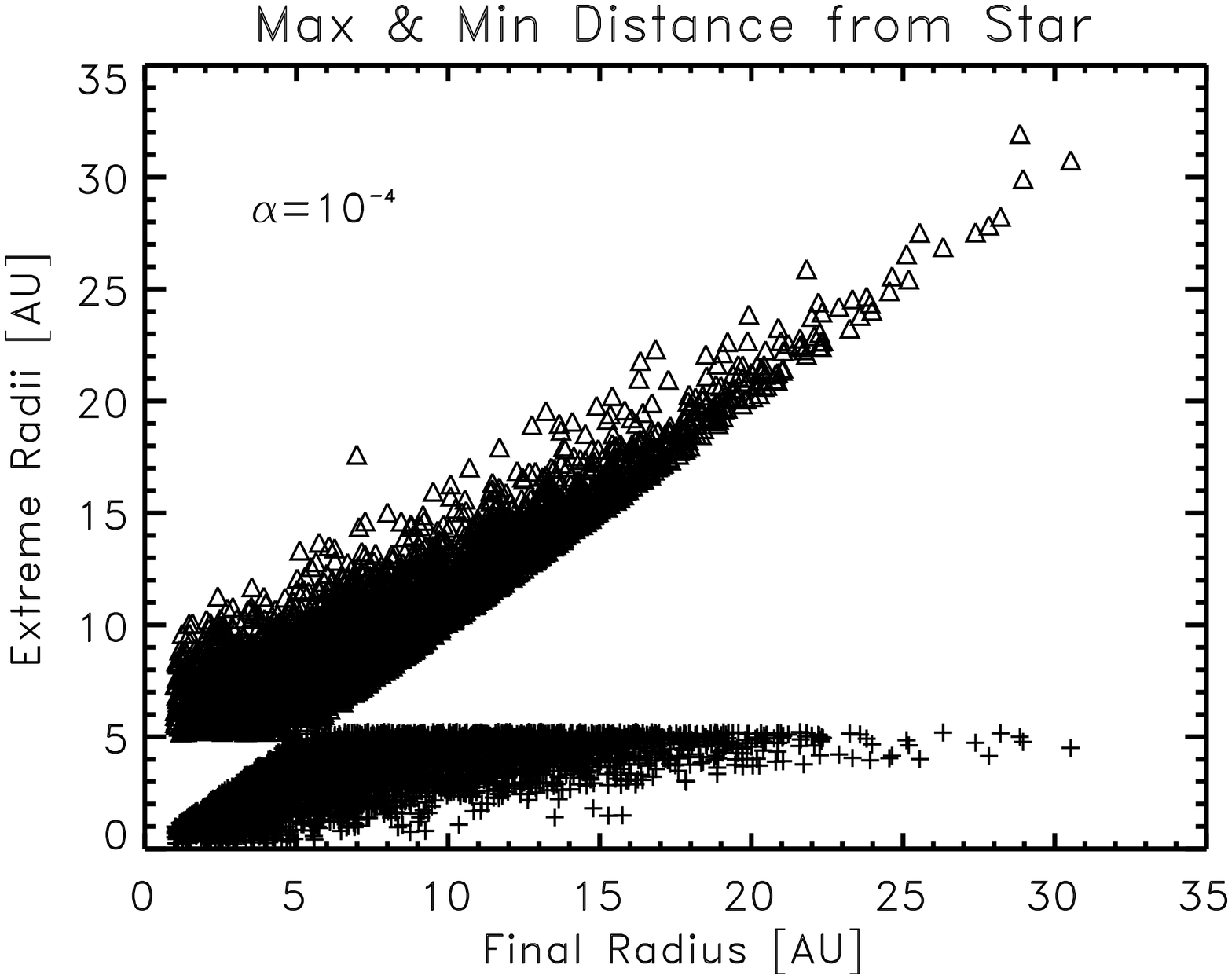}
\includegraphics[angle=90,width=3.1in]{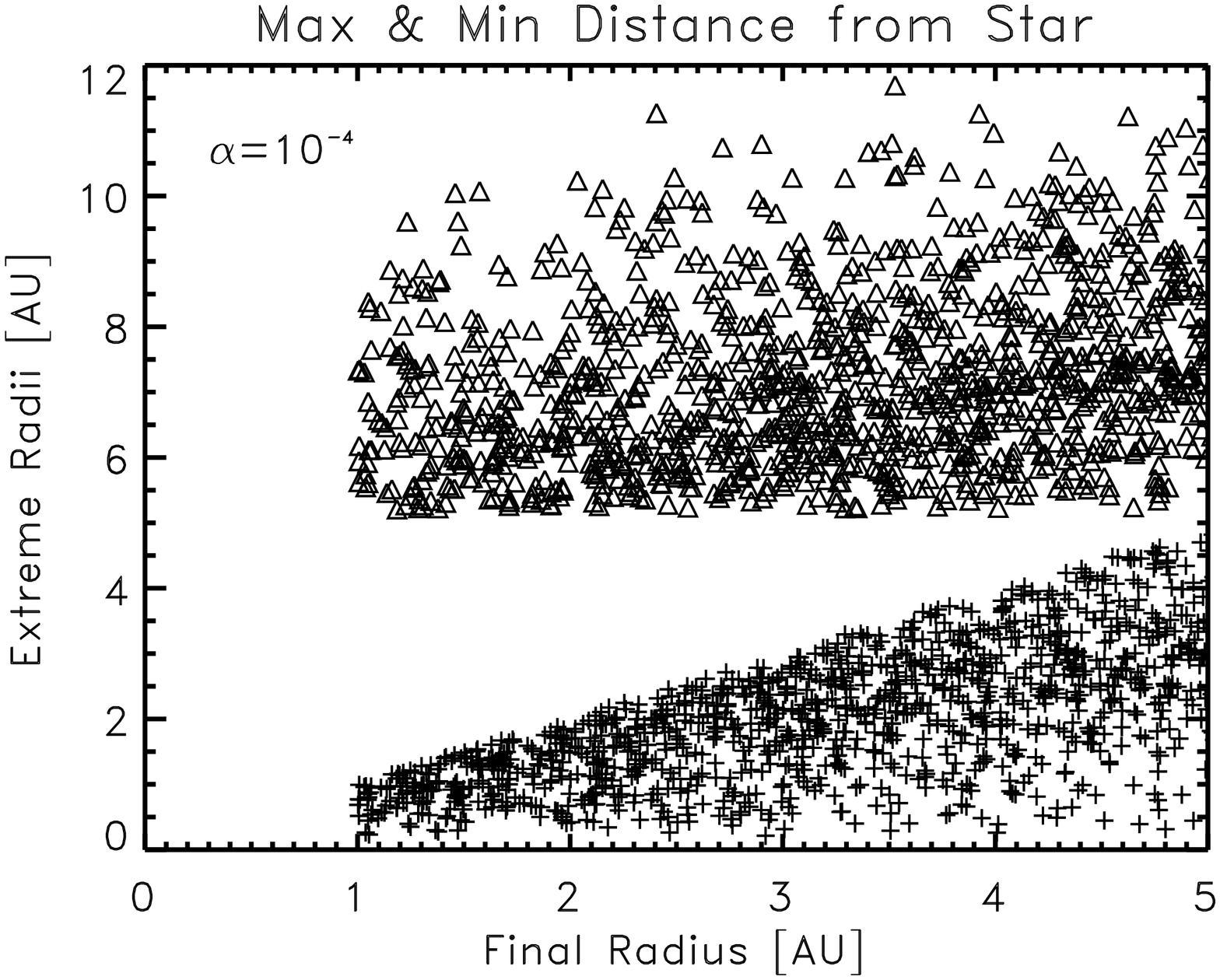}
\caption{Same as Figure 5, except the turbulence parameter is assumed to have a value of $\alpha$=10$^{-4}$.}
\end{center}
\end{figure}

\newpage
\begin{figure}
\begin{center}
\includegraphics[angle=90,width=3.1in]{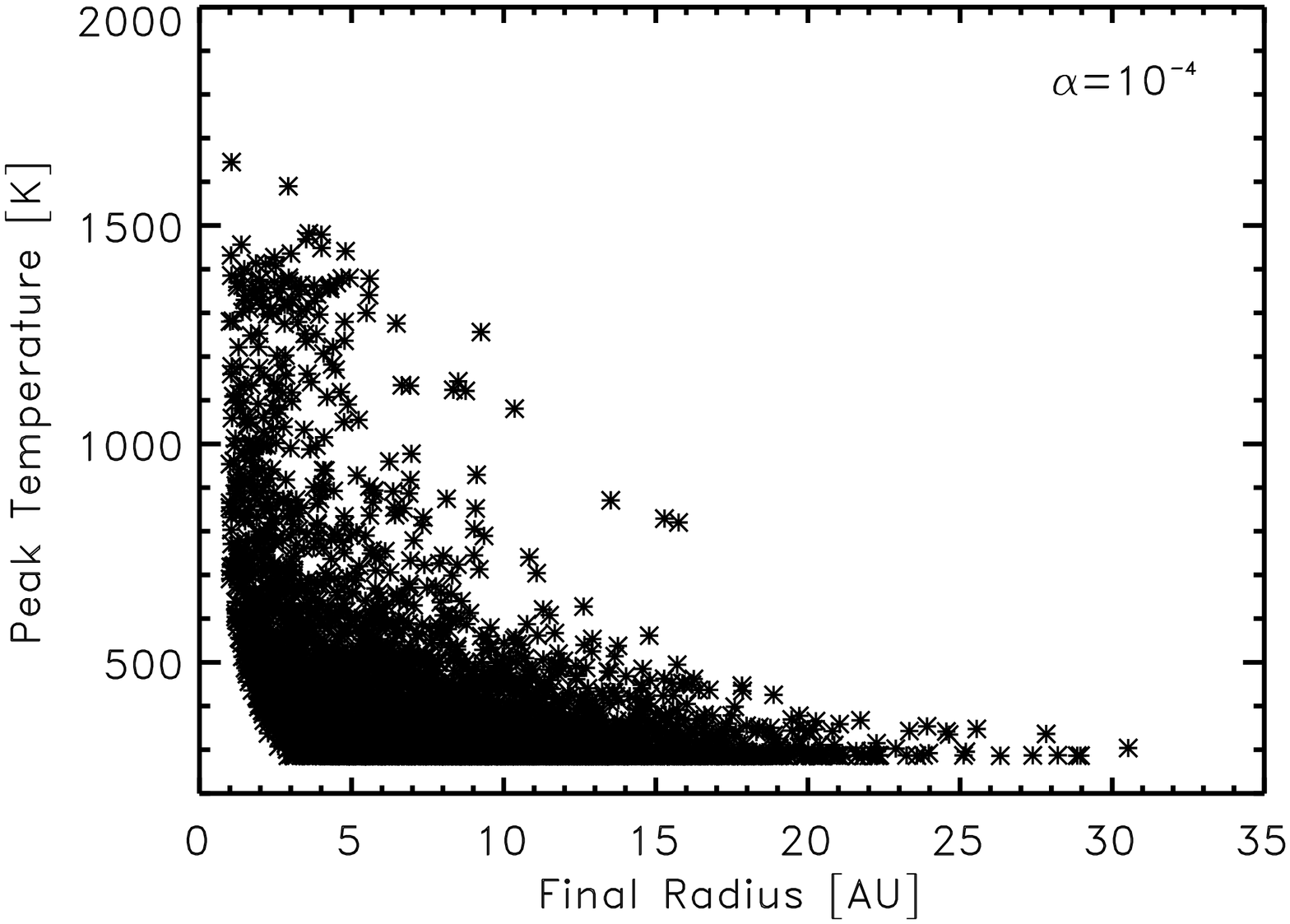}
\includegraphics[angle=90,width=3.1in]{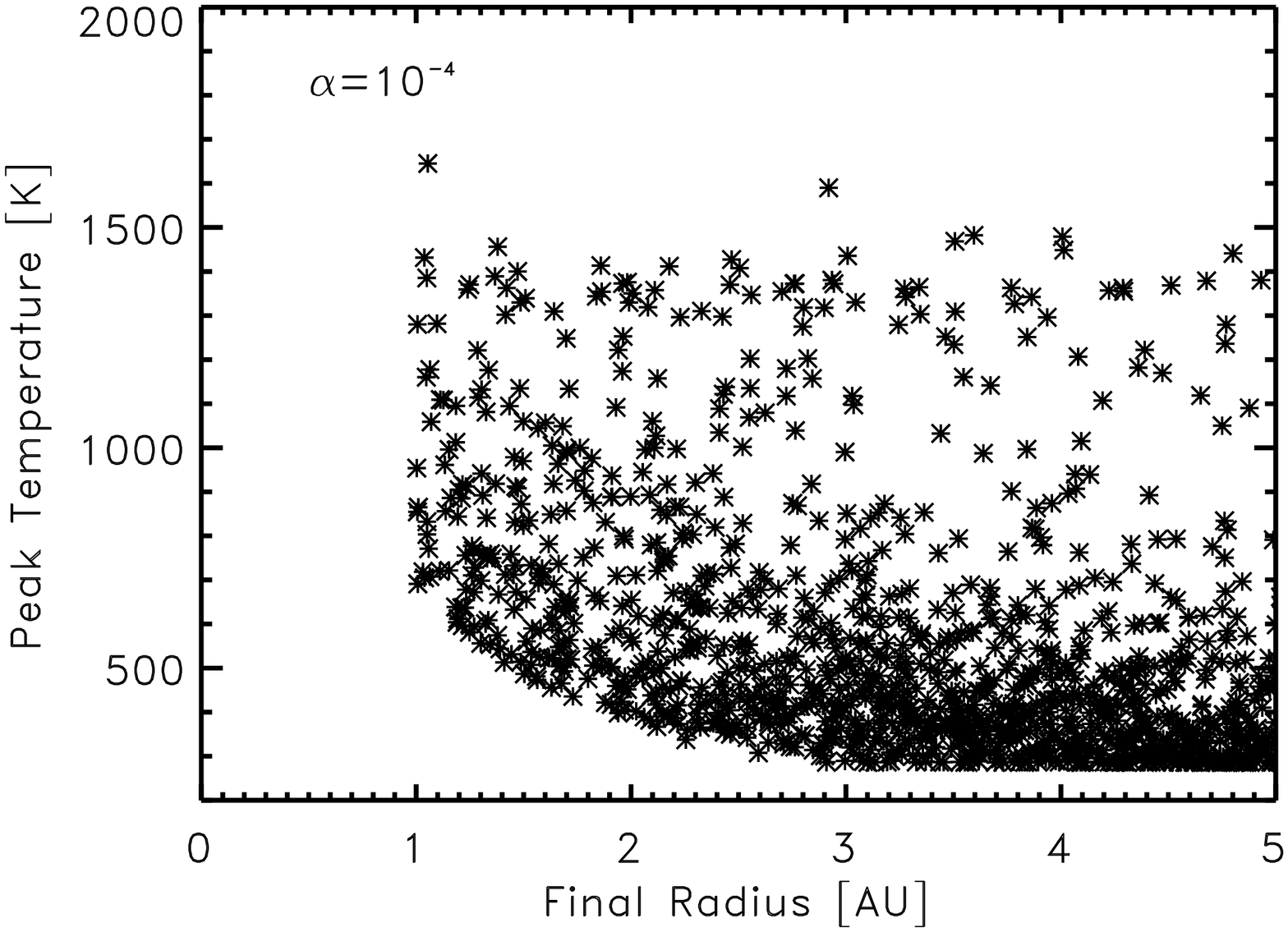}
\caption{Same as Figure 5, except the turbulence parameter is assumed to have a value of $\alpha$=10$^{-4}$.}
\end{center}
\end{figure}

\newpage
\begin{figure}
\begin{center}
\includegraphics[angle=90,width=5in]{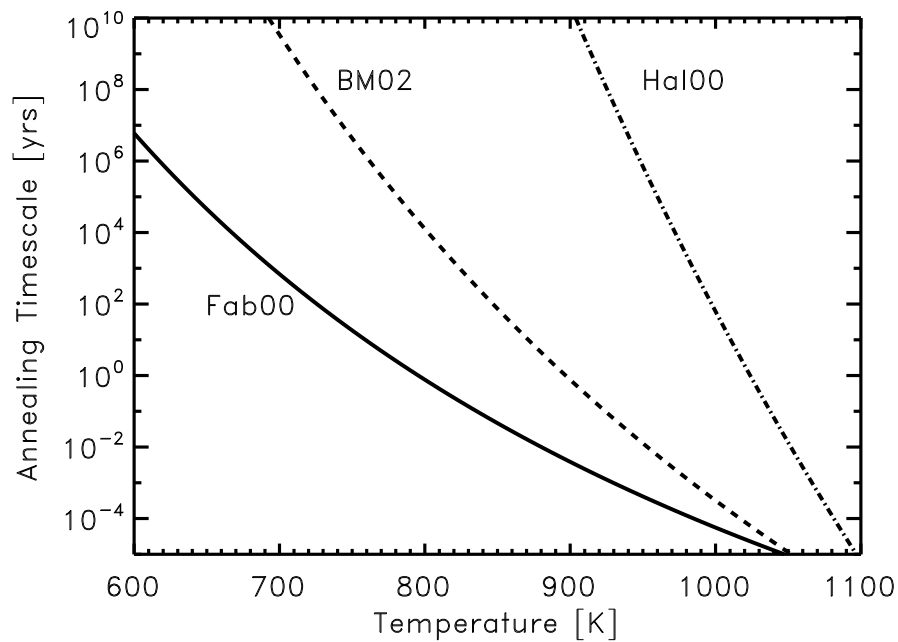}
\caption{Annealing timescales for the three annealing rates considered here: Mg smokes from \citet{fabian00} (Fab00: solid line), Mg glass as described by \citet{bockelee02} (BM02: dashed line), and Mg smokes as studied by \citet{hallenbeck00} (Hal00:dash-dot line).}
\end{center}
\end{figure}

\newpage
\begin{figure}
\begin{center}
\includegraphics[angle=90,width=3.1in]{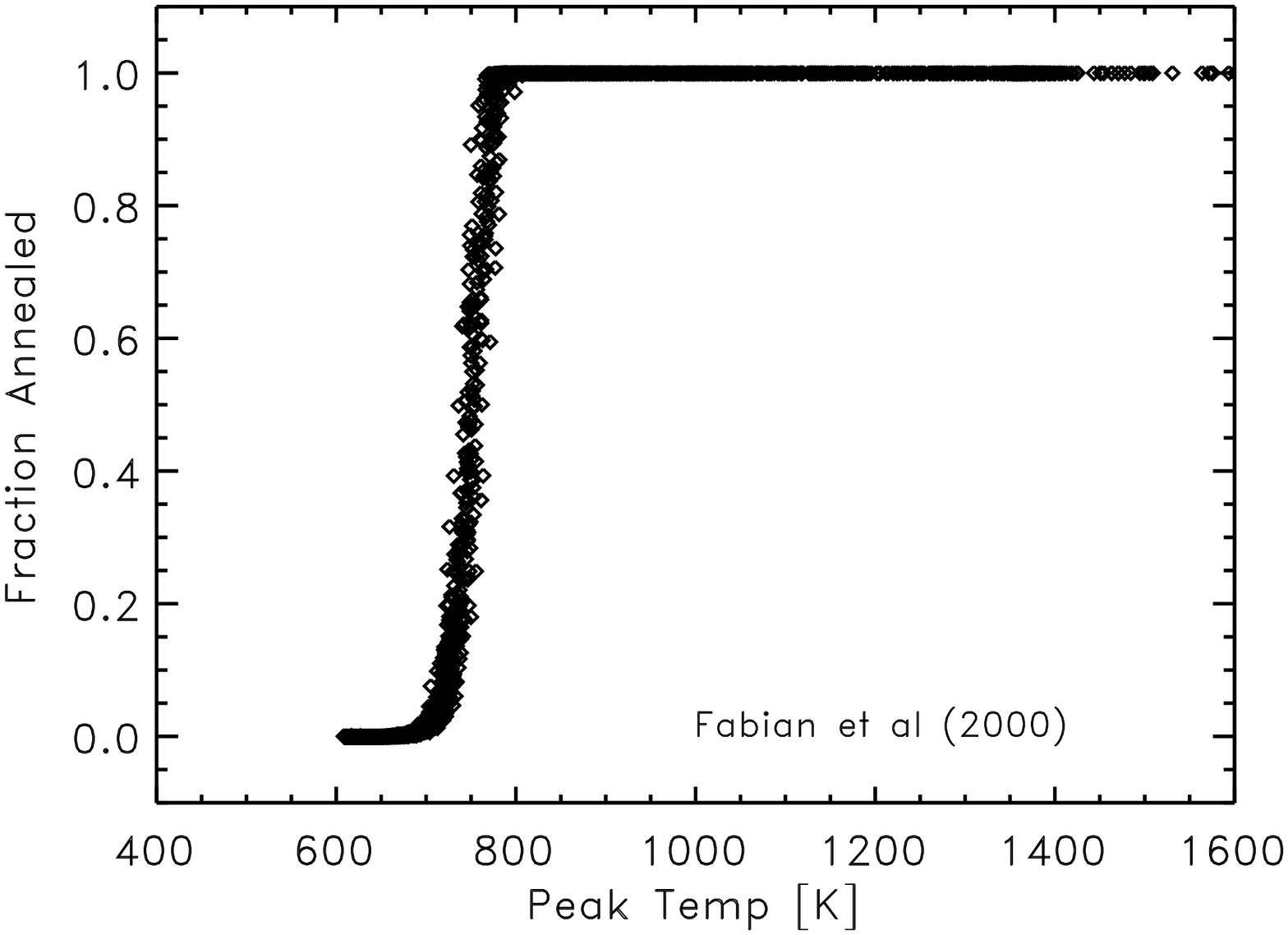}
\includegraphics[angle=90,width=3.1in]{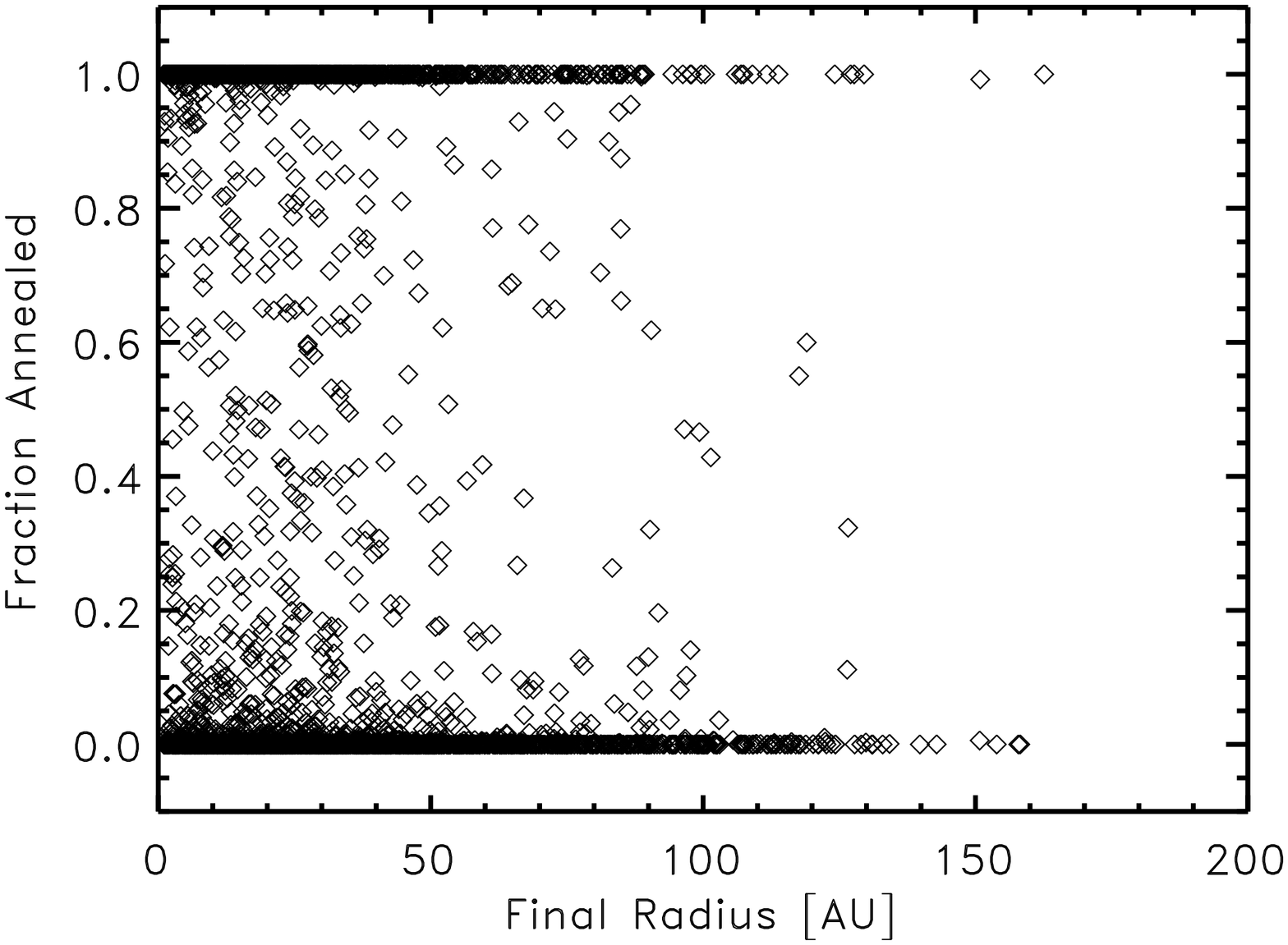}
\caption{\emph{Left:} The fraction of each grain that became crystalline (annealed) plotted as a function of the peak temperature seen by the grains, using the paths for those particles in the evolving disk described within the text and the \citet{fabian00} annealing law.  \emph{Right:} The fraction of each grain that became crystalline plotted versus its final location in the disk at the end of the simulation.}
\end{center}
\end{figure}

\newpage
\begin{figure}
\begin{center}
\includegraphics[angle=90,width=3.1in]{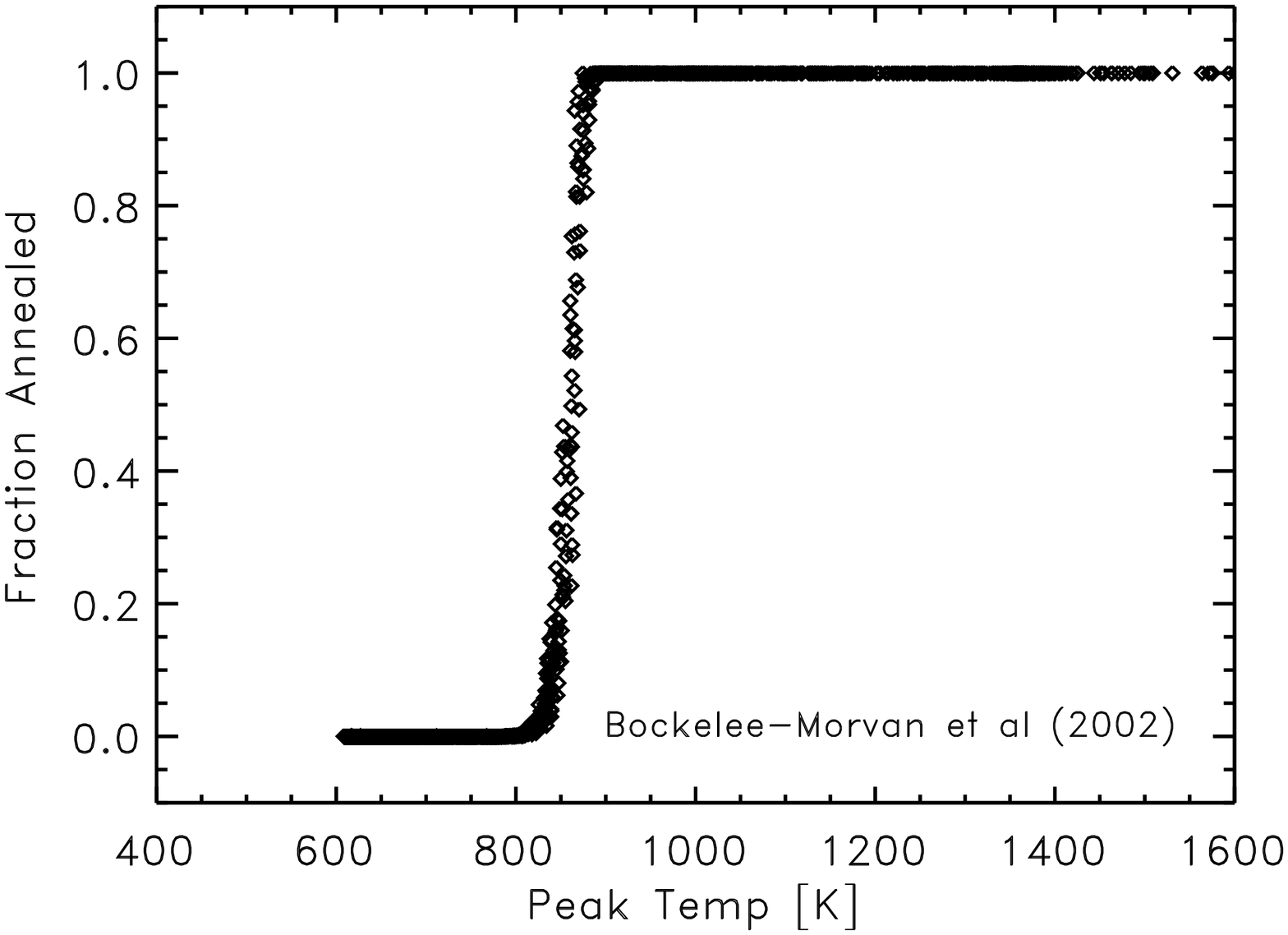}
\includegraphics[angle=90,width=3.1in]{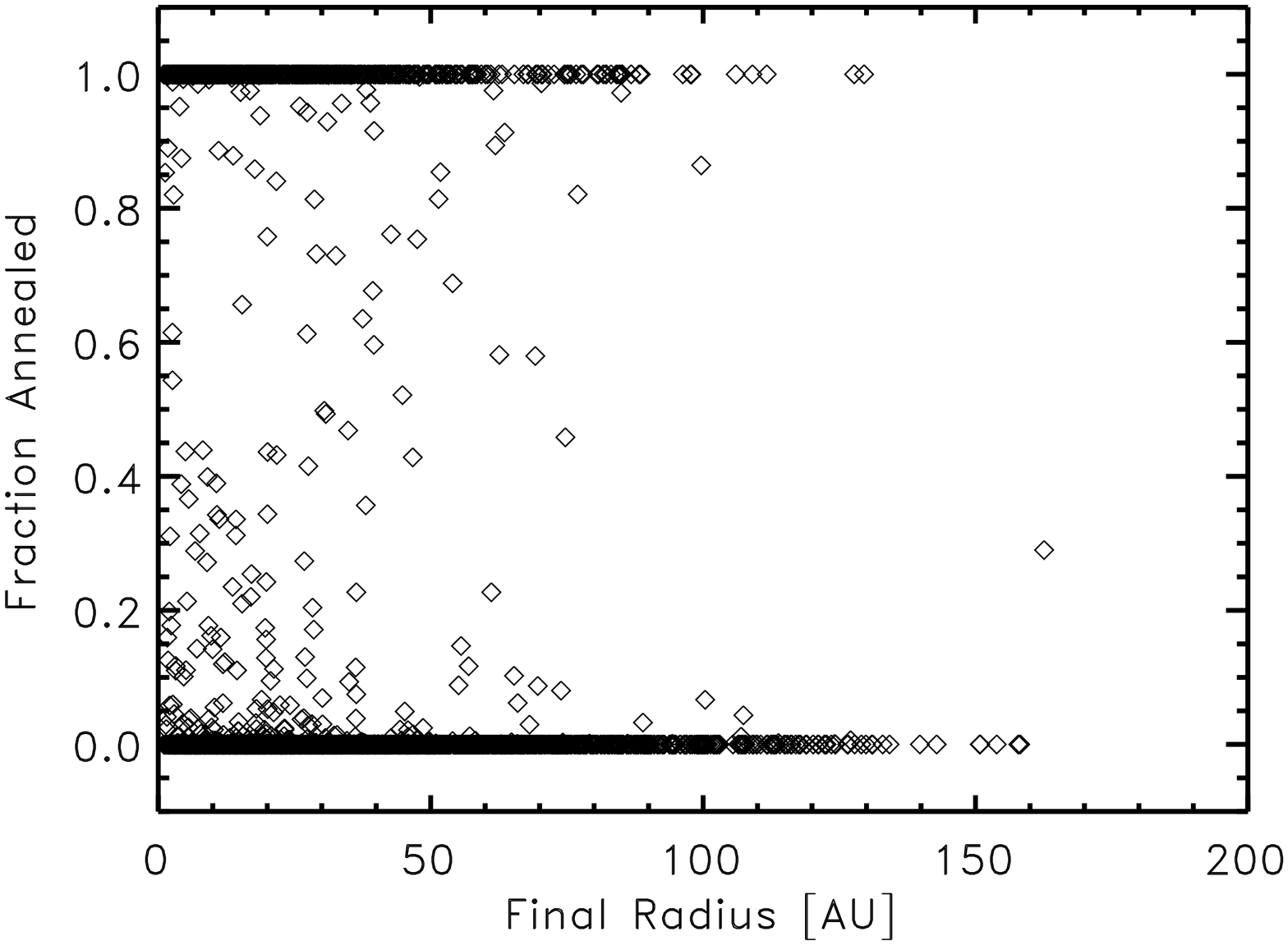}
\caption{Same as Figure 10, except using the \citet{bockelee02} annealing law.}
\end{center}
\end{figure}

\newpage
\begin{figure}
\begin{center}
\includegraphics[angle=90,width=3.1in]{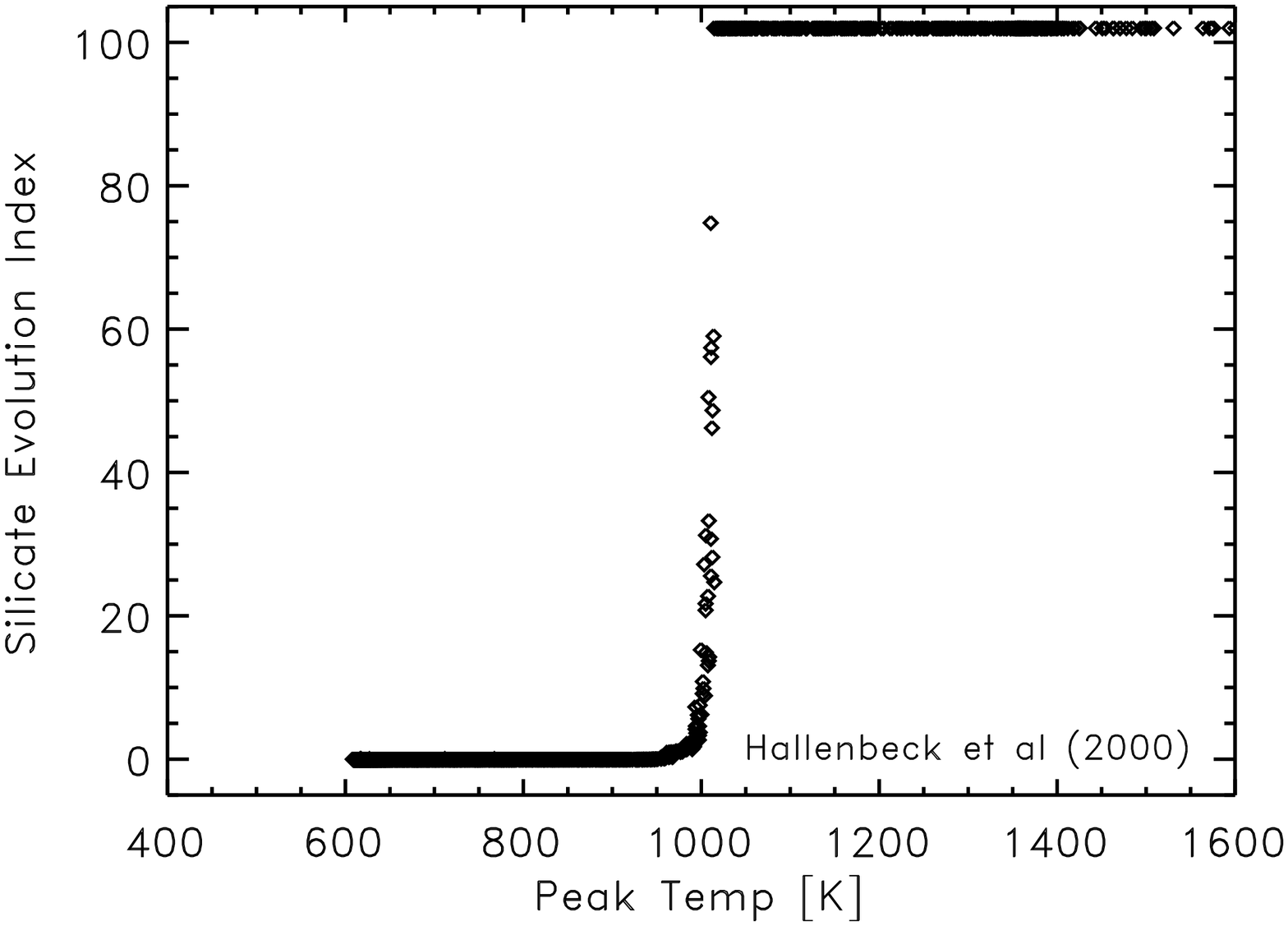}
\includegraphics[angle=90,width=3.1in]{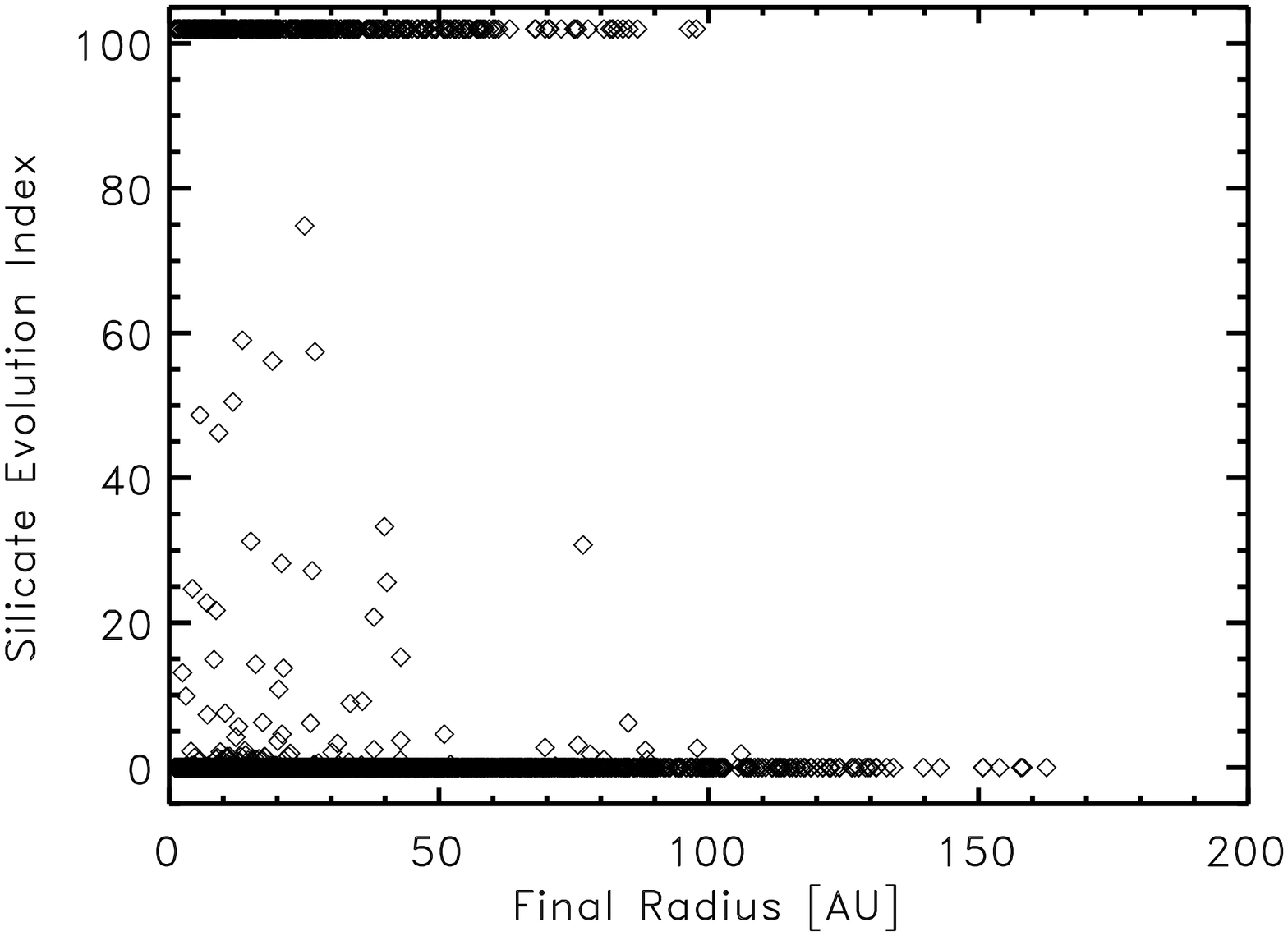}
\caption{Same as Figure 11, except using the Silicate Evolution Index of \citet{hallenbeck00}.}
\end{center}
\end{figure}

\newpage
\begin{figure}
\begin{center}
\includegraphics[angle=90,width=3.1in]{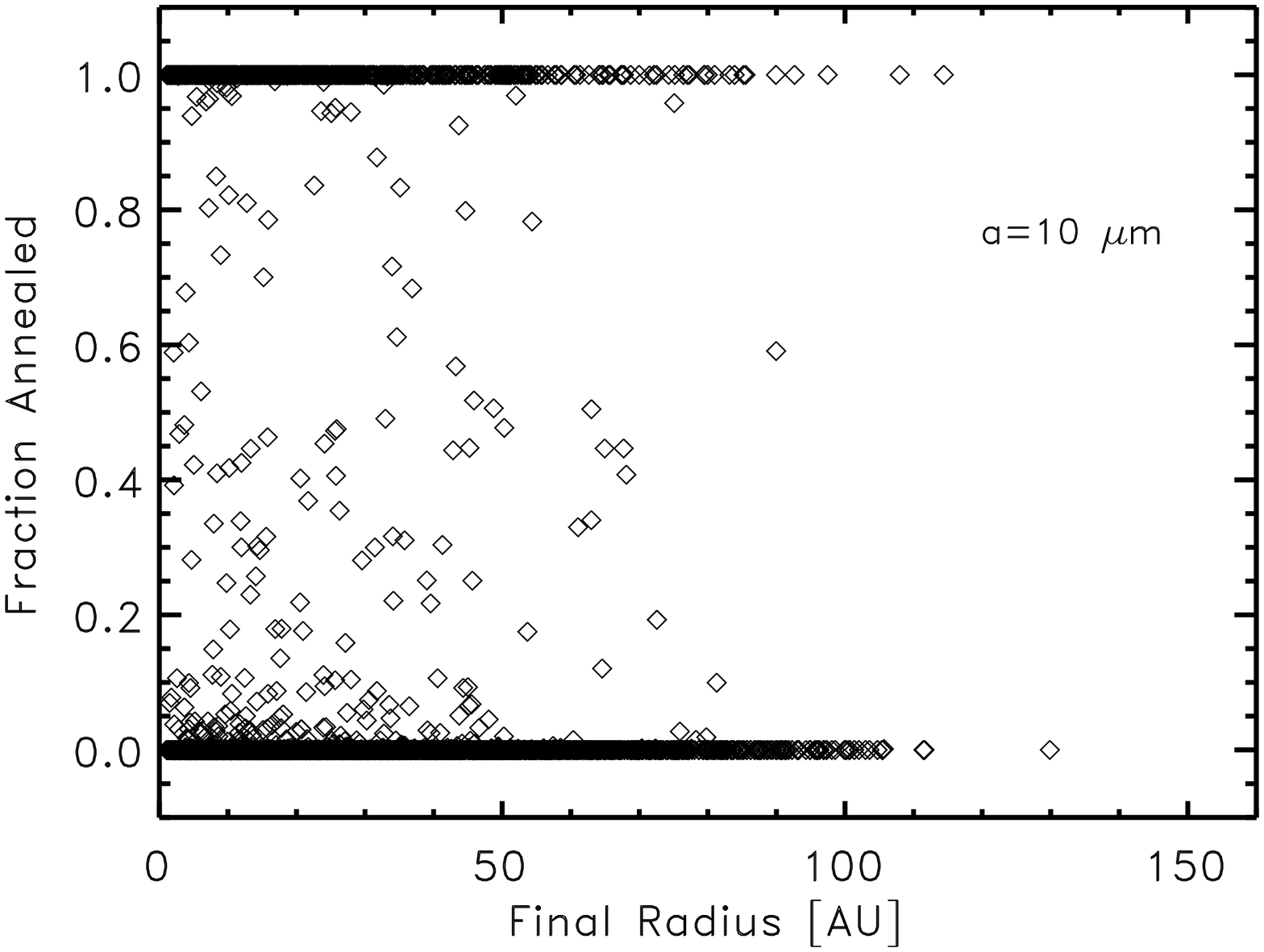}\\
\includegraphics[angle=90,width=3.1in]{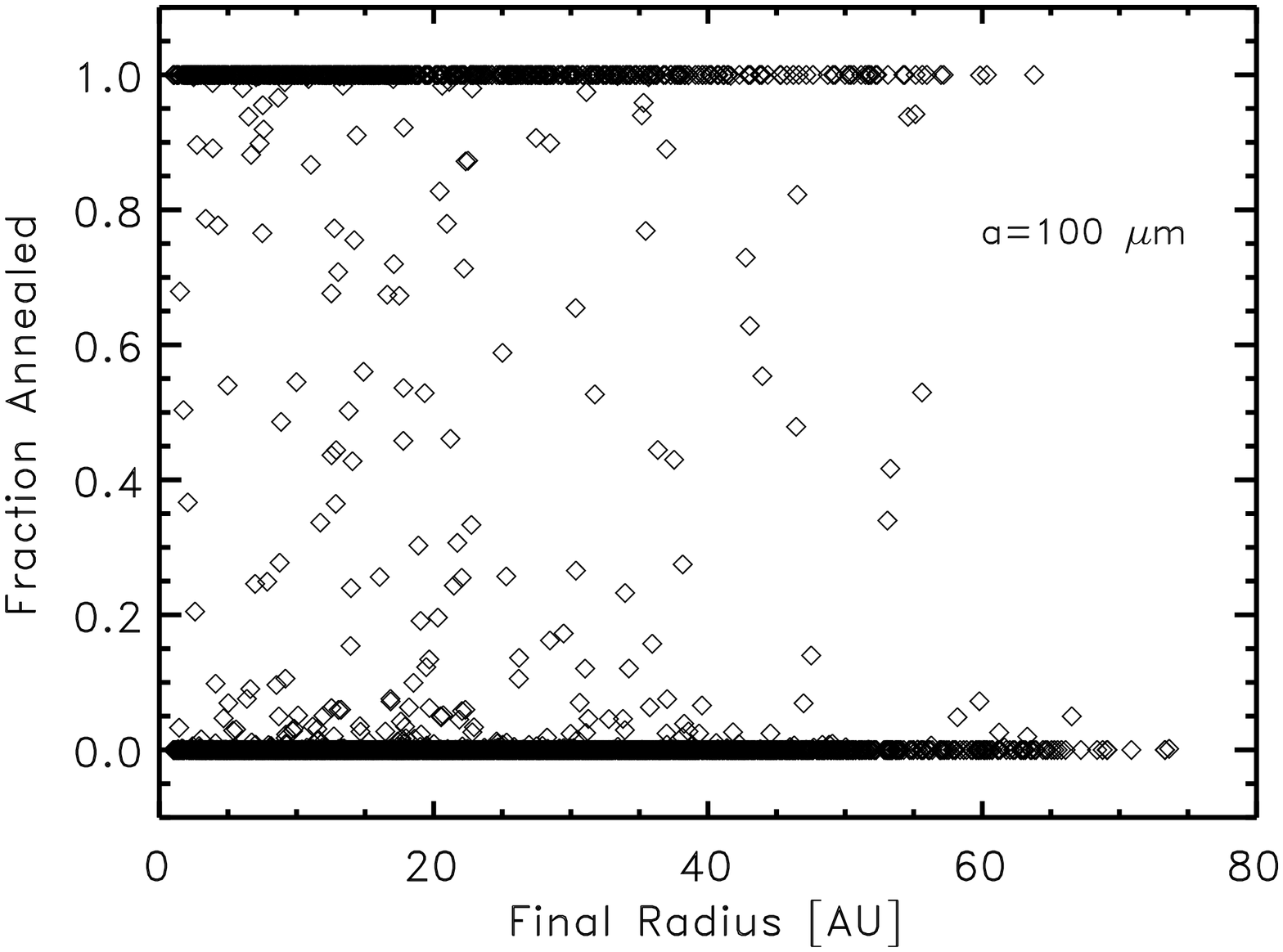}\\
\includegraphics[angle=90,width=3.1in]{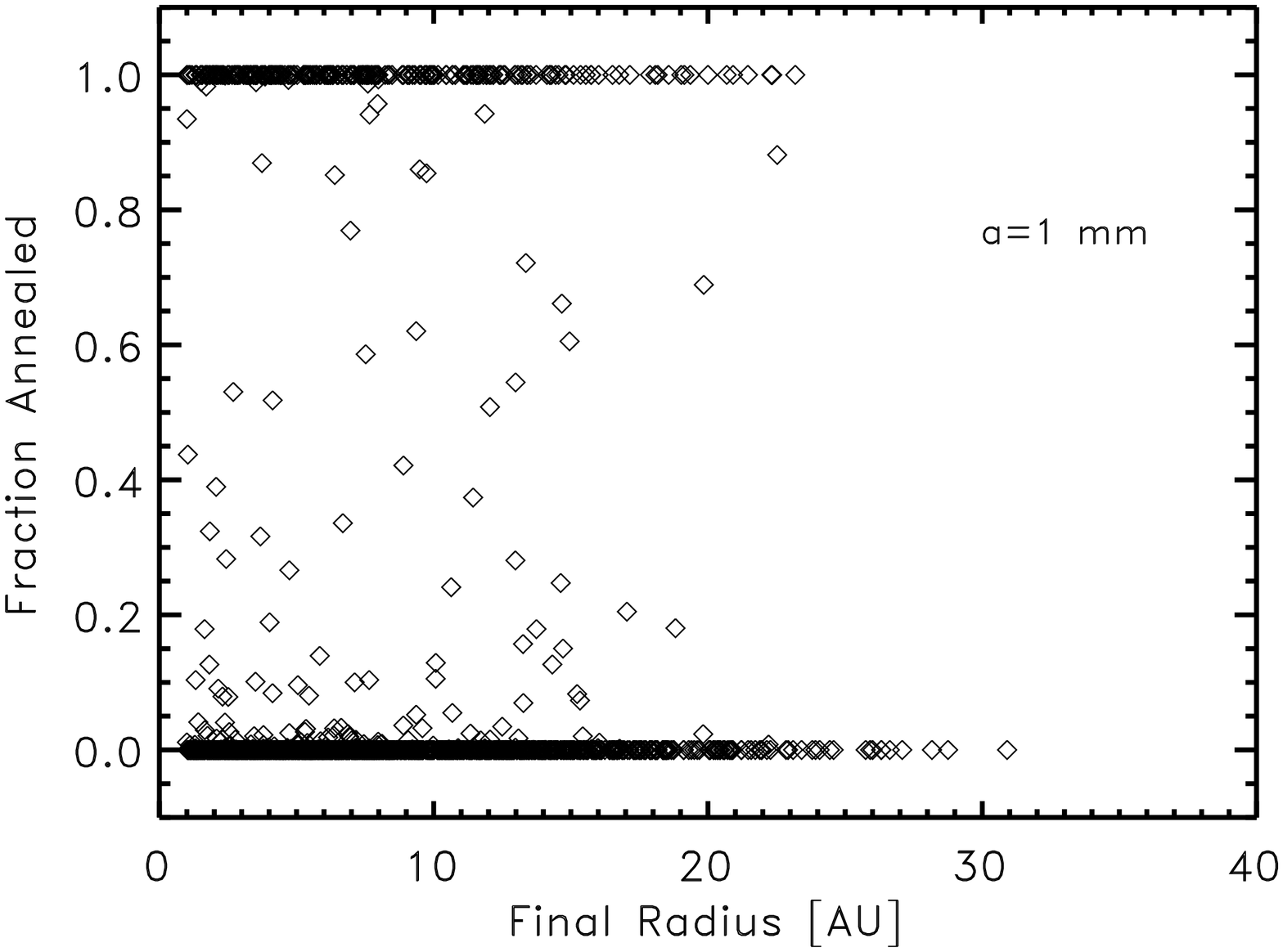}
\caption{The crystallinity distribution of particles of radius 10 $\mu$m, 100 $\mu$m, and 1 mm, run with the same disk model and assumptions described above, using the intermediate annealing law of \citet{bockelee02}.  Larger grains, regardless of their crystallinity, are confined to radial locations closer to the central star due to the effects of gas drag (note the different scales on the x-axis).  Despite the decrease in the survival frequency of the larger particles, the fraction of grains that are crystalline among the survivors is the same for all sizes considered here.}
\end{center}
\end{figure}

\newpage
\begin{figure}
\begin{center}
\includegraphics[angle=90,width=3.1in]{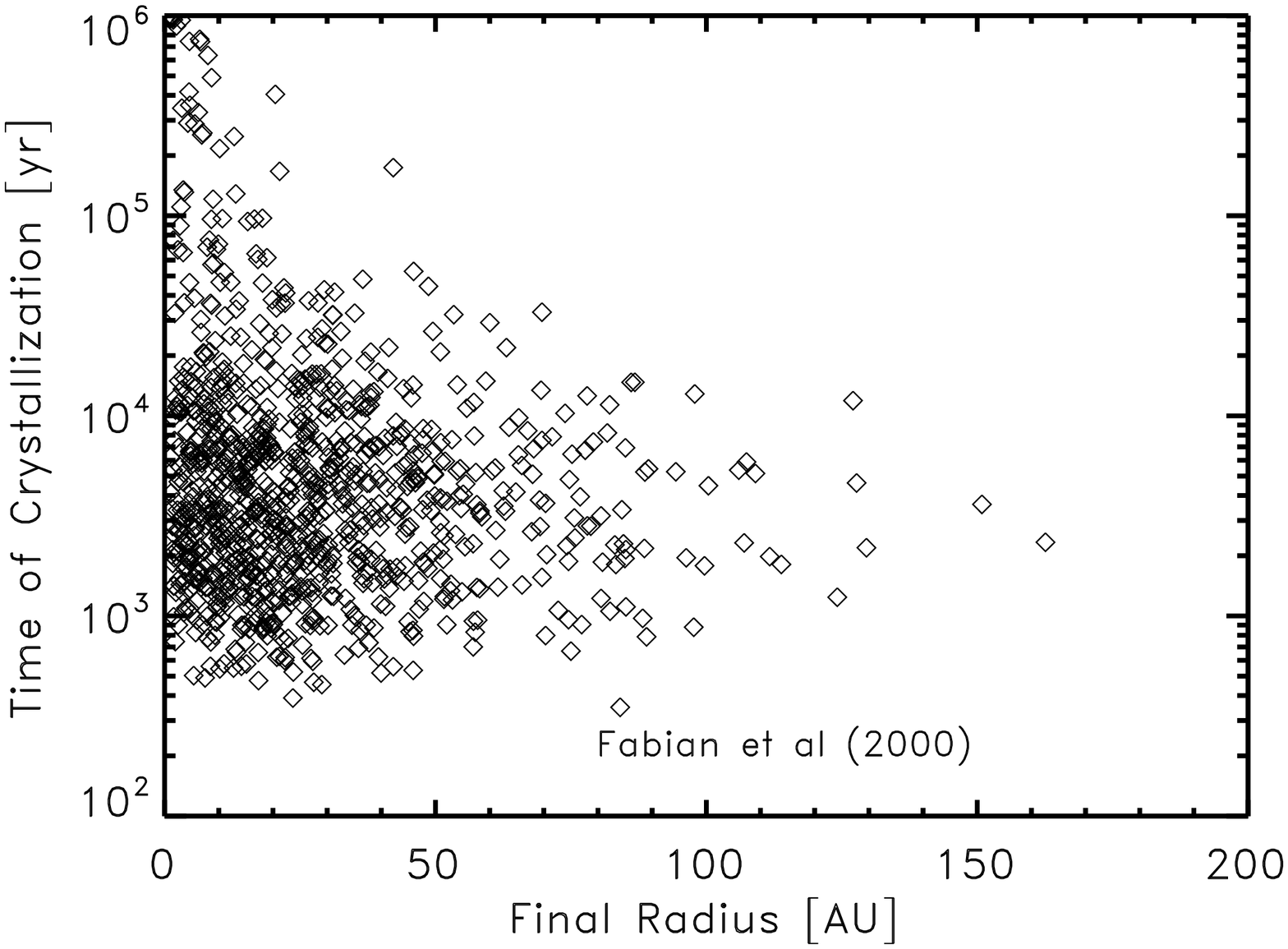}\\
\includegraphics[angle=90,width=3.1in]{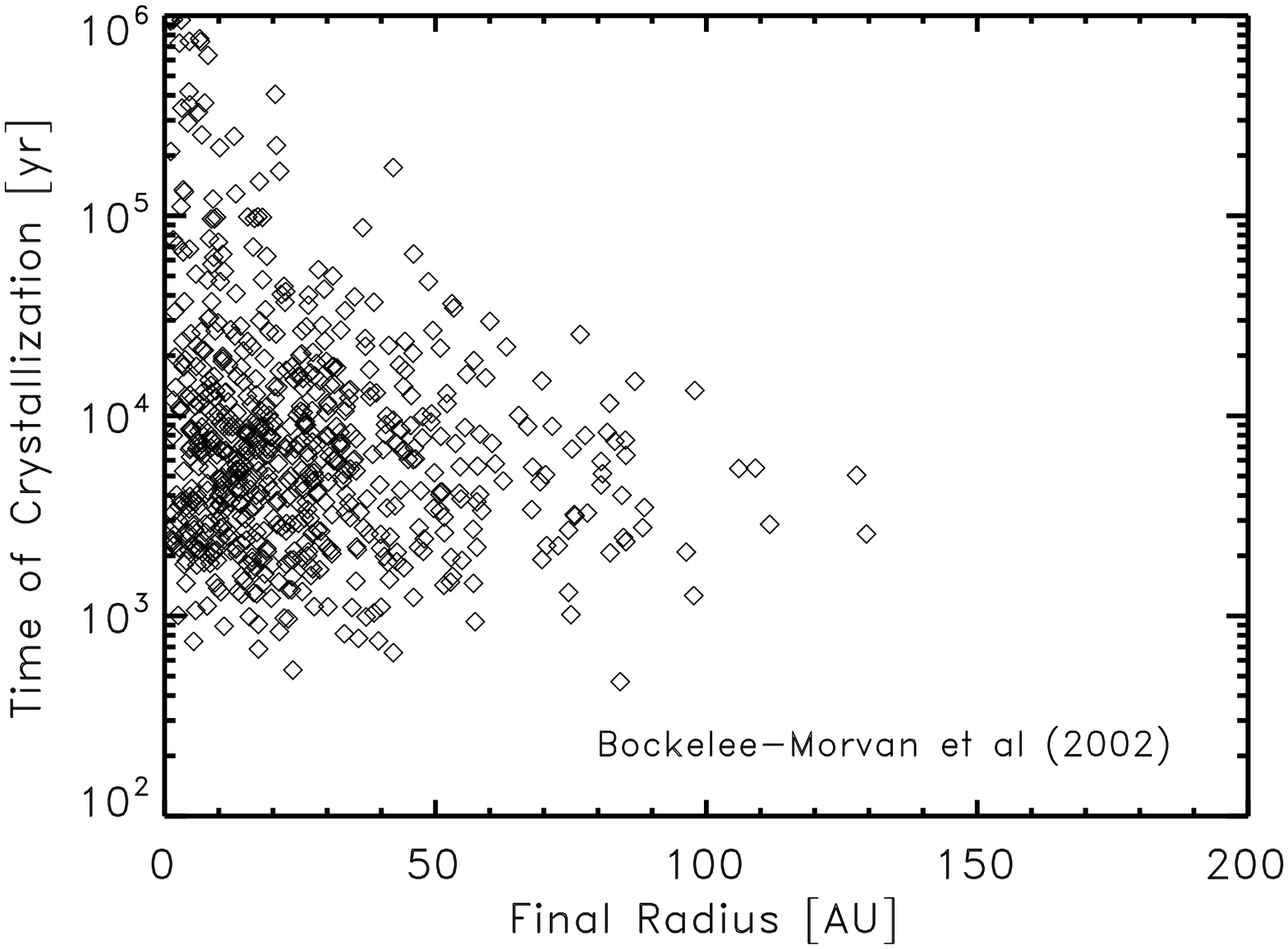}\\
\includegraphics[angle=90,width=3.1in]{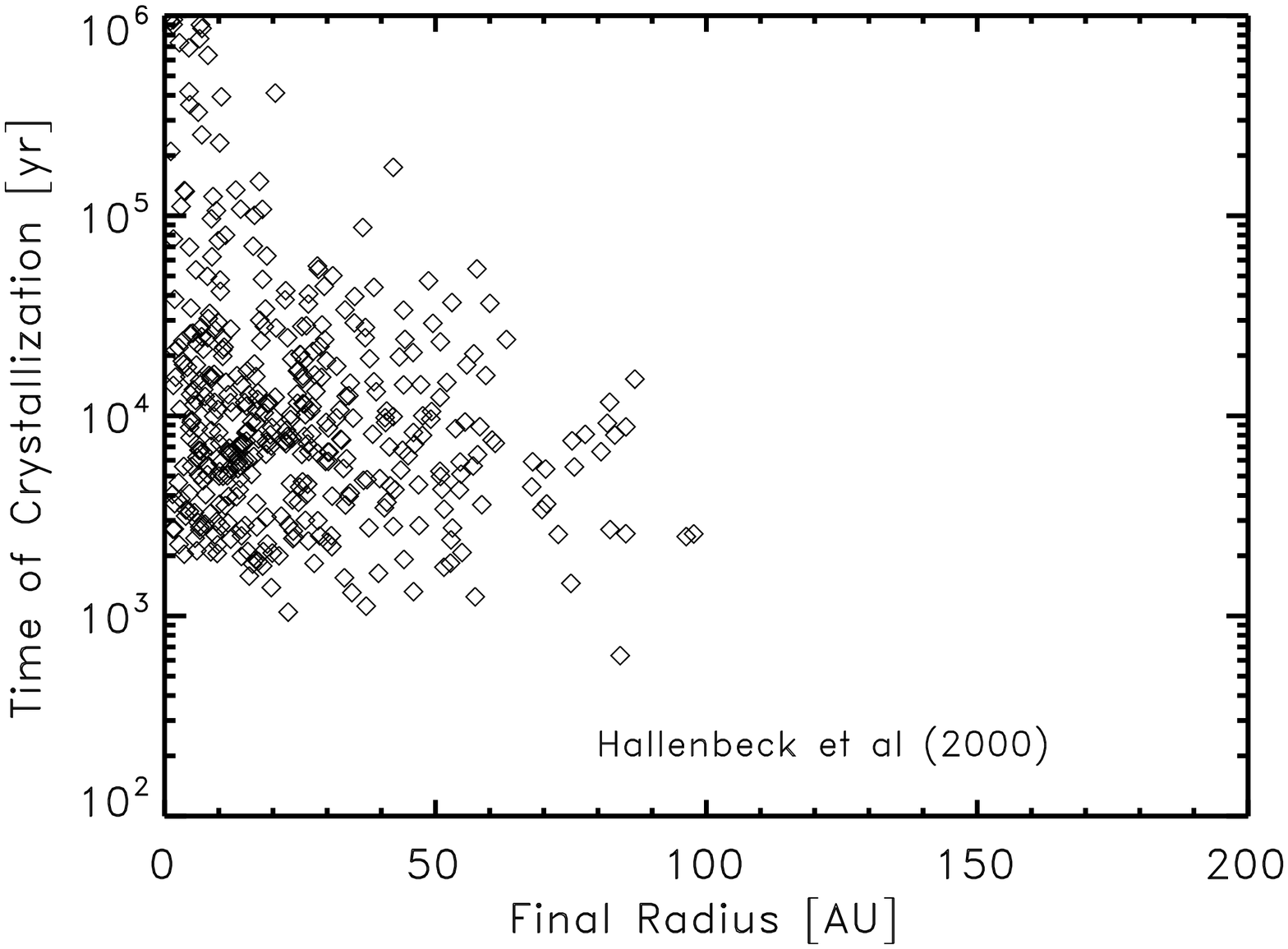}
\caption{The time that each grain became crystalline plotted versus the final location o the grain for the three annealing laws considered here.  Note that grains annealed late in disk history are limited to the inner disk, whereas those that make it to the outer disk come from a relatively small time window in the history of the disk.}
\end{center}
\end{figure}



\end{document}